\documentclass[prd, 10pt,aps,nofootinbib, floatfix, notitlepage,twocolumn,longbibliography]{revtex4-2}

\usepackage{amsmath,amsfonts,amssymb,enumitem}
\usepackage{mathrsfs}
\usepackage{graphicx}
\usepackage[english]{babel} 
\usepackage{url} 
\usepackage{xcolor}
\usepackage{lipsum}
\usepackage{bbold}
\usepackage{adjustbox}
\usepackage[utf8]{inputenc} 
\usepackage[colorlinks=true,citecolor=blue,urlcolor=blue]{hyperref}

\newcommand{\LCDM}{$\Lambda$CDM}
\newcommand{\tk}[1]{\textcolor{black}{#1}}
\newcommand{\kb}[1]{\textcolor{black}{ #1}}
\newcommand{\ede}{{\rm ede}}
\newcommand{\Neff}{{N_{\rm eff}}}
\begin{document}

\title{Thermal Friction as a Solution to the Hubble and Large-Scale Structure Tensions}
\author{Kim V. Berghaus$^{1}$}
\email{kim.berghaus@stonybrook.edu}
\author{Tanvi Karwal$^{2}$ }
\email{karwal@sas.upenn.edu}

\thanks{\\Both authors contributed equally.}

\affiliation{$^1$C.N. Yang Institute for Theoretical Physics, Stony Brook University, NY 11794, USA}
\affiliation{$^2$Center for Particle Cosmology, Department of Physics and Astronomy, University of Pennsylvania, Philadelphia, PA 19104, USA}

\begin{abstract}
Thermal friction offers a promising solution to the Hubble and the large-scale structure (LSS) tensions. 
This additional friction acts on a scalar field in the early universe and  extracts its energy density into dark radiation, the cumulative effect being similar to that of an early dark energy (EDE) scenario. 
The dark radiation automatically redshifts at the minimal necessary rate to improve the Hubble tension. 
On the other hand, the addition of extra radiation to the Universe can 
\kb{mitigate }the LSS tension. 
We explore this model in light of cosmic microwave background (CMB), baryon acoustic oscillation and \kb{type Ia} supernova data, including the SH0ES $H_0$ measurement and the Dark Energy Survey Y1 data release in our analysis. 
Our results indicate a preference for the regime where the scalar field converts to dark radiation at very high redshifts \kb{($z \gtrsim 10^5$)}, asymptoting effectively to an extra self-interacting radiation species rather than an EDE-like injection. 
In this limit, thermal friction can ease both the Hubble and the LSS tensions, but not resolve them. 
We find the source of this preference to be the incompatibility of the CMB data with the linear density perturbations of the dark radiation when injected at redshifts close to matter-radiation equality. 
\end{abstract}

\date{\today}

\maketitle

\section{Introduction}
\label{sec:intro}

The recent discoveries of growing discrepancies in cosmology, particularly the Hubble and large-scale structure tensions, may indicate new physics beyond the standard \LCDM\ concordance model. 
The Hubble tension is a mismatch between different estimations of the Hubble rate $H_0$ today \cite{Verde:2019ivm,Freedman:2017yms, Abdalla:2022yfr}. 
It has grown to the critical $5\sigma$-level between its two most precise constraints - one based on a \LCDM\ fit to the cosmic microwave background (CMB) as observed by Planck \cite{Aghanim:2018eyx} and the other from more direct measurements in the late universe using the distance-ladder approach by SH0ES \cite{Riess:2021jrx}. 
This mismatch is in fact echoed at a lower discrepancy level by several measurements \cite{DiValentino:2020zio,DiValentino:2020vnx}, with the early universe and a \LCDM\ model consistently finding a lower $H_0$ \cite{2013ApJS..208...19H,Addison:2017fdm,Aghanim:2018eyx,ACT:2020gnv,SPT-3G:2021eoc,Ivanov:2019pdj} than the late universe \cite{2012ApJ...758...24F,Riess:2016jrr,2018ApJ...855..136R,Pesce:2020xfe,Huang:2019yhh,Riess:2019cxk,Reid:2019tiq,Freedman:2019jwv,Yuan:2019npk,Freedman:2020dne,Riess:2020fzl,Gaia, deJaeger:2022lit,Cao:2022ugh, Riess:2021jrx}. 
A similar but milder tension is emerging in descriptions of the large-scale structure (LSS) of the Universe. 
Late-universe estimations of the amplitude $\sigma_8$ of matter fluctuations at a scale of $8 h^{-1}$Mpc are lower than those from a \LCDM\ fit to the early-universe CMB \cite{DES:2017myr, Hildebrandt:2018yau, HSC:2018mrq, KiDS:2020suj, DES:2021wwk,Nunes:2021ipq}. 
Ultimately, these tensions \kb{could} hint at an inconsistency between the early and late universes under \LCDM. 

Taking these tensions at face value, it has been challenging to postulate new-physics solutions \cite{Abdalla:2022yfr,DiValentino:2021izs,DiValentino:2020vvd,Schoneberg:2021qvd,Sunny_talk}. 
Late-universe modifications of \LCDM\ cosmology are constrained by supernovae and the consistency of BAO with the CMB \cite{Benevento:2020fev, Alestas:2021xes}, while early-universe modifications are constrained by precise measurements of the CMB \cite{Vagnozzi:2021gjh}. 
Nonetheless, two solutions demand further scrutiny. 
One is the introduction of extra free-streaming radiation to the Universe, usually in the form of additional massless neutrinos quantified by the effective number $\Neff$ of neutrino species \cite{Poulin:2018zxs, Kreisch:2019yzn, Aghanim:2018eyx,Riess:2016jrr}. 
This proposal is noteworthy because it can slightly ease both tensions, however it is unable to fit cosmological data well, in particular, it worsens the fit to the CMB.
Another notable solution is the addition of early dark energy (EDE) \cite{Karwal:2016vyq, Poulin:2018dzj, Poulin:2018cxd, Smith:2019ihp, Agrawal:2019lmo, Lin:2019qug, Sakstein:2019fmf, CarrilloGonzalez:2020oac, Niedermann:2019olb, Alexander:2019rsc, Allali:2021azp, Freese:2021rjq,Ye:2020btb,Braglia:2020bym,Gogoi:2020qif,Garcia:2020sjl,Ye:2021iwa}, a new component which behaves like a cosmological constant at early times, then dilutes as fast or faster than radiation, such that its impact on cosmology 
is localized in redshift. 
Although this solution resolves the Hubble tension, it exacerbates the LSS tension \cite{Ivanov:2020ril, Hill:2020osr, Smith:2020rxx, Murgia:2020ryi,Secco:2022kqg}.
Moreover, EDEs have been under scrutiny on the theory front
for relying on extremely fine-tuned scalar-field potentials \cite{Kamionkowski:2014zda,Poulin:2018dzj, Berghaus:2019cls, Gonzalez:2020fdy, Karwal:2021vpk}, and for not offering an explanation for why the scalar field becomes dynamic close to matter-radiation equality, a shortcoming that has been dubbed the ``why-then" problem \cite{Karwal:2021vpk, Sakstein:2019fmf}. 

In previous work \cite{Berghaus:2019cls}, 
we proposed thermal friction acting upon a scalar field as a solution to
the Hubble tension, combining the favorable characteristics of both the EDE and the extra-radiation solutions. 
This model circumvents the need for fine-tuned potentials and holds promise for providing a good fit to data, addressing a criticism each of the two models it consolidates. 

In this scenario, a scalar field experiences thermal friction $\Upsilon$ in addition to Hubble friction. 
This extracts the energy density of the decaying scalar into dark radiation. 
The dark radiation automatically redshifts at the minimum required rate for EDE scenarios, obviating the need for finely-tuned scalar field potentials, a crucial development from a model-building perspective.
The cumulative scalar field and dark radiation energy densities then provide an EDE-like energy injection into the early universe, with the added species diluting like radiation, similar to $\Neff$, potentially alleviating the Hubble and the LSS tensions simultaneously.  

Such models have been considered in the context of both inflation \cite{Berera:1995ie, Berera:1995wh, Berera:1999ws, Berera:1998px, Berera:2008ar, Bastero-Gil:2016qru, Berghaus:2019whh} and late-time dark energy \cite{Graham:2019bfu, Berghaus:2020ekh} due to their desirable model-building properties as well as unique predictions for observations. 
Hence, besides having favourable characteristics to address both tensions, this model can also provide explanations for two other eras in cosmic history, with a cascading family of axions experiencing thermal friction spread out over redshift. 

In this paper, we perform a detailed study of the cosmological implications of thermal friction in the context of the Hubble tension. 
We present our model in Sec.~\ref{sec:thermal friction} and derive the perturbations and initial conditions of this theory in synchronous gauge.
We then explore parameter constraints for this model with methodology, data sets and priors described in Sec.~\ref{sec:data_method}. 
We present these constraints for various dataset combinations in Sec.~\ref{sec:results}, \kb{with (\ref{sec:results_H0}) and without (\ref{sec:results_baseline})} a SH0ES $H_0$ prior and including DES data (\ref{sec:results_LLS}). 
We find that the preferred injection redshift for our model asymptotes to high $z$, \kb{where DA EDE mimics an extra self-interacting dark radiation solution.} 
In Sec.~\ref{sec:results_higher_z_c}, we demonstrate that expanding our parameter space will not yield phenomenologically different results and explore the origins of this preference in Sec.~\ref{sec:results_no_EDE_solution}.
Finally, we summarise our findings and conclude in Sec.~\ref{sec:conclusions}.

\section{Thermal Friction}
\label{sec:thermal friction}

Couplings between scalar fields and light degrees of freedom are a natural extension of minimal scalar field models and have long been considered in other cosmological contexts
\cite{Berera:1995ie, Berera:1995wh, Berera:1999ws, Berera:1998px, Berera:2008ar, Bastero-Gil:2016qru}. 
In \cite{Berghaus:2019cls}, we introduced a coupling between an axion field $\phi$ and a dark non-Abelian gauge group [SU(2)] which induced a thermal friction $\Upsilon$ in the equation of motion of the scalar field.
In this minimal model, the axion field $\phi$ injects its potential energy into dark radiation comprised of dark gauge bosons, instead of converting it into its own kinetic energy. 
The dark radiation efficiently self-interacts, maintaining a (dark) thermal environment,\footnote{
This requirement is trivially fulfilled when the thermal friction $\Upsilon$ exceeds the Hubble rate.} 
and suppressing shear perturbations. 
This is in contrast with extra radiation in the form of neutrino species which have non-negligible shear perturbations due to their free-streaming \cite{Ma:1995ey,Kreisch:2019yzn,Poulin:2018zxs}. 

Inspired by this dissipative axion (DA) model, we explore thermal friction in the context of the Hubble tension. 
We start by defining the stress-energy tensors for the DA sector. 
We decompose its energy content into the dark radiation component (dr) and the scalar field ($\phi$) component:
\begin{equation}
    T^{\mu \nu}_{\text{DA}} = T^{\mu \nu}_{\phi} + T^{\mu \nu}_{\text{dr}} \,.
\end{equation}
Then $T^{\mu \nu}_{\text{DA}}$ is conserved and does not transfer energy to the usual $\Lambda$CDM components. 
However, there is energy transfer between the $\phi$-field and the dark radiation, $-\nabla_{\mu} T^{\mu \nu}_{\phi} =  \nabla_{\mu} T^{\mu \nu}_{\text{dr}}  $\,, which we quantify by \cite{Bastero-Gil:2014raa} 
\begin{equation} \label{stress}
    - \nabla_{\mu} T^{\mu \nu}_{\phi} = g^{\nu \alpha} \left( -\Upsilon {v_{\text{dr}} ^\mu} \partial_{\mu} \phi \,  \partial_{\alpha} \phi \right) \,.
\end{equation}
Here $v_{\text{dr}}^{\mu} = \frac{dx^{\mu}}{dt}$ is the 4-velocity of the dark radiation with respect to proper time $t$. 

In a self-consistent model, the macroscopic friction coefficient $\Upsilon(\rho_{\text{dr}})$, which allows for energy-momentum exchange between the scalar field and the dark radiation, emerges from the coupling between the scalar field and the light fields which make up the radiation in the theory.  
We treat $\Upsilon$ as a constant for the in-depth analysis in this paper.
The full microphysical DA model has temperature-dependent thermal friction ($\Upsilon(\rho_{\text{dr}}) \propto \rho^{\frac{3}{4}}_{\text{dr}}$), which we do not consider here.\footnote{
As we shortly explain, our theory is valid in the regime where the scalar field is overdamped by the additional friction $\Upsilon$ and undergoes no oscillations. 
Temperature-dependence introduces multi-dimensional non-linear boundaries between the underdamped and overdamped regime, making a systematic exploration of the overdamped regime non-trivial. 
}
Thus, when referring to DA EDE throughout this paper, we imply a constant friction $\Upsilon$, unless otherwise indicated. 
We briefly comment on consequences of temperature dependence in \ref{subsec:thermal friction perturbation}, and derive the perturbations for a general temperature dependence in appendix \ref{sec:derivations}.

\subsection{Background evolution}

Using the stress-energy tensor to derive the evolution equation of the scalar field and the dark radiation (see Appendix \ref{sec:derivations} for details), we find 
\begin{eqnarray} \label{eom}
    \phi'' + \left(2 \mathcal{H}  +   a \Upsilon \right) {\phi}' + a^2 V_\phi &=& 0  \,,
    \nonumber \\ 
    \rho'_{\text{dr}} + 4 \mathcal{H} \rho_{\text{dr}}  &=& \frac{\Upsilon}{a} {\phi'}^2  \,.
    \label{eq:eqs_of_motion}
\end{eqnarray}
Here primes indicate derivatives with respect to conformal time $\tau$, and $\mathcal{H} \equiv \frac{a'}{a}$, where $a$ is the scale factor.
We take the potential to be quadratic $V(\phi) = \frac{1}{2}m^2 \phi^2$, \kb{where $m$ denotes the mass of the scalar field,}  and $\rho_{\text{dr}}$ denotes the dark radiation energy density. 
We focus on a simple quadratic potential since the alleviation of fine-tuning was our initial motivation to explore thermal friction, but our results hold for more complex potentials. 
\tk{In the regime preferred by data, all terms in which the dynamics of the scalar field enter become negligible, thus making our results independent of the choice of scalar field potential.}\footnote{\kb{Scalar-field potentials $V(\phi) = \lambda_n \phi^n$ with $n>2$ are challenging to obtain from a UV-complete theory, as dominant terms with $n = 2$ are also generated in the underlying theory (see \cite{Gonzalez:2020fdy}, Eq.~2, for instance). 
A quadratic potential on the other hand arises naturally. 
For example, for the axion-like particle we consider here it could be obtained through an explicit symmetry breaking term in the UV-theory, making the axion-like particle a pseudo-Nambu Goldstone boson with a small mass term. }}

We restrict our analysis to the overdamped regime in which the constant friction $\Upsilon \geq m$. 
In this regime, thermal friction solves the fine-tuning of EDE-potentials. 
In the standard ultra-light-axion scalar-field EDE scenario, the scalar field is frozen until $\frac{\mathcal{H}}{a} \sim m$, at which point it rapidly begins to oscillate. 
In the DA EDE model, the thermal friction dominates over Hubble friction ($a\Upsilon \gg \mathcal{H}$), and the scalar field never oscillates. 

The time dependence of the overdamped scalar field is well described by 
\begin{equation} \label{osc}
    \phi(\tau) \approx \phi_0 e^{-\frac{m^2}{\Upsilon} \int^\tau_0 a(\tau')d\tau'} \,.
\end{equation}
Approximating $ \int^\tau_0 a(\tau')d\tau' \sim \frac{a}{\mathcal{H}}$\,, 
we estimate that the thermal friction system decays away when 
\begin{equation}
    \frac{a\Upsilon}{m^2} \lesssim \mathcal{H}(z) \,.
\end{equation}
This timescale also coincides with the peak energy density of the dark radiation, approximately at the redshift $z_c$ of the maximal fractional contribution $f_\ede$ of DA EDE to the total cosmic energy budget, 
\begin{equation} \label{eq:fede}
    f_{\text{ede}}(z_c) = \frac{\rho_{\text{ede}}(z_c)}{\rho_{\text{tot}}(z_c)} \,,    
\end{equation}
where 
\begin{equation}
    \rho_{\text{\text{ede}}}(z) = \rho_{\phi}(z) +\rho_{\text{dr}}(z) \,,
\end{equation}
and $\rho_{\text{tot}}$ is the sum over all components present in the early universe. 
Hence, $\frac{\Upsilon}{m^2}$ maps directly onto the phenomenological EDE parameter $z_c$.

\begin{figure}
    \centering
    \includegraphics[width=0.49\textwidth]{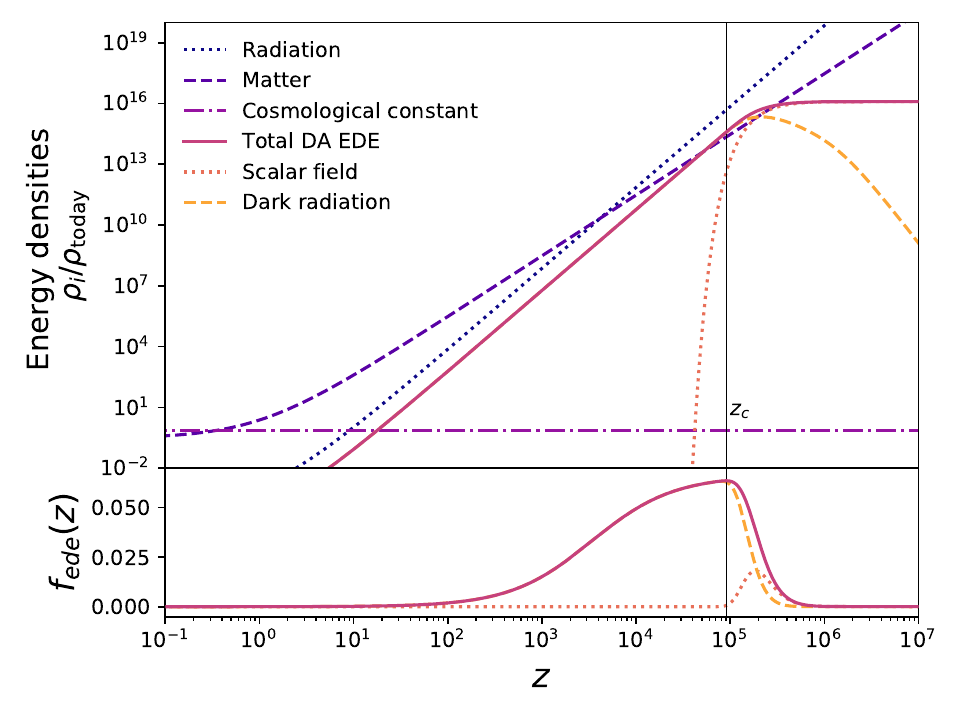}
    \caption{
    The energy densities $\rho_i/\rho_{\rm today}$ of various components in units of the critical density today are shown here against redshift $z$, for the best-fit DA EDE cosmology when fitting to \textit{baseline}+$H_0$ (defined in Sec. \ref{sec:datasets}). 
    These include radiation (dotted curve), matter (dashed), the cosmological constant (dot-dashed) and DA EDE (solid). 
    We further split DA EDE into its two components, a scalar field component (dashed) dominant in the early universe that rapidly vanishes near the critical redshift $z_c$ (marked by the vertical line); and dark radiation (dotted), which grows in the early universe, dominates near $z_c$ and then dilutes $\propto (1+z)^4$. 
    In the lower panel, we show the fractional energy density in DA EDE and its components. 
    }
    \label{fig:bg_ede}
\end{figure}

If the decay timescale is close to matter-radiation equality at $z_{\rm eq}$, thermal friction reproduces a modified EDE scenario \cite{Berghaus:2019cls}. 
If $z_c \gg z_{eq}$, this system asymptotes to an extra-radiation solution with self-interacting dark radiation. 
We refer to these as EDE-like and extra-radiation regimes respectively, and show the latter in Fig. \ref{fig:bg_ede}, which corresponds to our best fit DA EDE cosmology.  

\subsection{Linear Perturbation dynamics}
\label{subsec:thermal friction perturbation}

We derive the density and velocity perturbation equations of the scalar field as well as the dark radiation fluid in synchronous gauge using Eq.~\eqref{stress} and find 
\begin{equation} 
    \label{eq:phipt}
    \delta \phi'' + 2 \mathcal{H} \delta \phi' + \left(k^2 +a^2 V''(\phi) \right) \delta \phi  = -\frac{h' \phi'}{2} -  \Upsilon a  \delta \phi' \,,
\end{equation}
\begin{equation} 
    \label{eq:rhopt1}
    -\frac{2h'}{3} -\frac{4}{3} \theta_{\text{dr}} + 2\frac{\Upsilon}{a \rho_{\text{dr}}} \delta \phi' \phi' = \delta'_{\text{dr}} +  \frac{\Upsilon{\phi'}^2}{a \rho_{\text{dr}}}  \delta_{\text{dr}} \,,
\end{equation}
\begin{equation} 
    \label{eq:rhopt2}
    {\theta'}_{\text{dr}} + \frac{\Upsilon}{a \rho_{\text{dr}}}  \phi'^2 {\theta}_{\text{dr}}   =   +  \frac{k^2}{4} \delta_{\text{dr}}  + k^2 \frac{3\Upsilon}{4a \rho_{\text{dr}}} \phi' \delta \phi \,.
\end{equation}   
Here, we have defined $\delta_{\text{dr}} \equiv \frac{\delta \rho_{\text{dr}}}{\rho_{\text{dr}}}$, and $\theta_{\text{dr}} \equiv i k^i v_i$.
Note that in the $\Upsilon \to 0$ limit, these equations reduce to those of a regular scalar field \cite{Smith:2019ihp}, and strongly self-interacting radiation in synchronous gauge \cite{Ma:1995ey}. 
See Appendix \ref{sec:derivations} for details on the derivation.  

The scalar field perturbations now have an additional friction sink term $\Upsilon a \delta \phi'$ that extracts energy from the scalar field perturbations, in parallel with the modifications to its background evolution. 
The fluid equations are modified by both a source term $2\frac{\Upsilon}{a \rho_{\text{dr}}} \delta \phi' \phi'$, and a sink term $\frac{\Upsilon{\phi'}^2}{a \rho_{\text{dr}}}  \delta_{\text{dr}}$ in Eq.~\eqref{eq:rhopt1}. 
The sink term is due to the modified continuity equation in Eq. \eqref{eom}, which enters when rewriting $\delta \rho_{\text{dr}}'$ in terms of $\delta_{\text{dr}}'$. 
Similarly, the velocity perturbations have an addition source $k^2 \frac{3\Upsilon}{4a \rho_{\text{dr}}} \phi' \delta \phi$, and sink term $\frac{\Upsilon}{a \rho_{\text{dr}}}  \phi'^2 {\theta}_{\text{dr}}$.

Practically, we find that the scalar-field perturbations have a negligible impact on CMB spectra.
As the scalar field exponentially decays for $z < z_c$,
it has inconsequential contribution to the energy budget at $z_*$ when the CMB is emitted. 
Accordingly, the source terms in the fluid equations that are \kb{proportional to} $ \delta \phi$ also quickly become sub-dominant. 

We distinguish two regimes for terms impacting CMB spectra. 
For very early injection times, $z_c \gg z_{\text{eq}}$, the system quickly approaches one in which all thermal friction terms are unimportant. 
In effect, this regime is identical to a universe that has always included self-interacting dark radiation, as opposed to having it injected at some redshift. 
The impact on CMB spectra is then dominated by the addition of such dark radiation. 
For injection around $z_c \sim z_{\rm eq}$, although $\phi$ and $\delta \phi$ quickly vanish, $\phi'$ remains sizeable enough such that the sink term $\frac{\Upsilon{\phi'}^2}{a \rho_{\text{dr}}}  \delta_{\text{dr}}$ dominates the evolution in equation \eqref{eq:rhopt1}. 
This leads to a suppression of the overdensities of the dark radiation fluid. 
In essence, the continuous sourcing of dark radiation acts to smooth its anisotropies, differentiating the evolution of its perturbations from those of other fluids which are dictated by metric perturbations. 
We examine how this dynamic affects the CMB in Sec. \ref{sec:results_no_EDE_solution}. 

The case of temperature-dependent friction (which we do not explore in this work) also sees this suppression of dark radiation density perturbations in the EDE-like regime. 
The term responsible for the smoothing in this scenario is $\left(1 - \frac{n}{4}\right)\frac{\Upsilon{\phi'}^2}{a \rho_{\text{dr}}}  \delta_{\text{dr}}$\,, for temperature-dependent friction $\Upsilon \propto \rho_{\rm dr}^{\frac{n}{4}}$. 
As attractor initial conditions only exist for $n < 4$ \cite{Berghaus:2019whh},
the sink term can at most decrease by a factor of 4. 
Thus, even for temperature dependent friction we expect a smoothing of the dark radiation anisotropies which differentiates its perturbation evolution from those of the other fluids.  

In the extra-radiation regime, with $z_c \gg z_{eq}$, we do not expect qualitative differences between constant and temperature-dependent friction, since all friction-related terms become sub-dominant at redshifts the CMB is sensitive to.

\subsection{Initial Conditions}

Our system has attractor initial conditions at the background level. 
At very early times, when the axion field is effectively frozen, we can assume $\phi'' \to 0$ and 
\begin{equation}
    \phi' \simeq \frac{a m^2 \phi}{2\mathcal{H}+a \Upsilon} \,.
\end{equation}
Plugging this estimate into Eq.~\eqref{eom} and neglecting $\rho'_{\text{dr}}$, we find 
\begin{equation}
    \rho_{\text{dr}} \simeq  \frac{\Upsilon a m^4 \phi^2 }{4\mathcal{H}(2\mathcal{H}+a\Upsilon)^2} \,.
\end{equation}
Neglecting $\rho'_{dr}$ when estimating the initial amount of dark radiation is justified as this system maintains a quasi-steady-state temperature on timescales $\sim (a\Upsilon)^{-1}$ (or $(2\mathcal{H})^{-1}$, whichever is shorter at the time initial conditions are set). Note that due to the attractor initial conditions the system quickly approaches these conditions even when starting with more or less dark radiation.
The initial value $\phi_i$ is an input parameter that controls how much new physics we are injecting. This corresponds to the amount of EDE-like energy, or when the scalar field decays away very early how much extra dark radiation is injected.
We assume adiabatic initial conditions for the DA EDE perturbations:
\begin{align}
    \delta_{\phi} &= 0 \,, \\
    \delta_{\text{dr}} &= \frac{3}{4}\delta_{\gamma} \,, \,\,{\rm and}  \\
    \theta_{\text{dr}} &= \theta_{\gamma} \,. 
\end{align}

\section{Datasets and Methodology}
\label{sec:data_method}

We add the above DA EDE cosmology to the Boltzmann code \verb|CLASS| \cite{Blas:2011rf}, and run Markov chain Monte Carlo (MCMC) simulations using \verb|Cobaya| \cite{Torrado:2020dgo} to obtain parameter posterior distributions, defining convergence using the Gelman-Rubin criterion \cite{Gelman:1992zz}, $R - 1 < 0.05$. 
We use \verb|GetDist| \cite{Lewis:2019xzd} for analyzing output. 
Lastly, to obtain best-fit parameter values, we utilize the \verb|BOBYQA| likelihood maximization code \cite{powell2009bobyqa, cartis2018escaping, cartis2019improving}. 

\subsection{Data}
\label{sec:datasets}

For parameter constraints, we consider the
\tk{standard datasets that EDE investigations employ} \cite{Poulin:2018cxd, Smith:2019ihp,Niedermann:2019olb,Niedermann:2020dwg,Agrawal:2019lmo,Lin:2019qug,Ye:2020btb}: 
\begin{enumerate}
    \item the Planck 2018 CMB high-$\ell$ (TTTEEE), low-$\ell$ (lowl + lowE) \cite{Aghanim:2018eyx, Planck:2019nip} and lensing measurements \cite{Planck:2018lbu};
    \item BAO measurements from BOSS DR12 at redshifts $z = 0.38, 0.51,\, \text{and } \, 0.614$ \cite{BOSS:2016wmc}, SDSS Main Galaxy Sample at z = 0.15 \cite{Ross:2014qpa}, and 6dFGS at z = 0.106 \cite{Beutler:2011hx}; and
    \item the Pantheon Supernovae (SN) sample \cite{Pan-STARRS1:2017jku}.\footnote{We note that while a newer version of this dataset has been released, the corresponding likelihood is not yet out. Moreover, we expect minimal changes to our constraints from improvements in SNe, as DA EDE all but vanishes at the redshifts where SNe provide constraints.
    The primary purpose of SNe data here is to eliminate the possibility of resolving the Hubble tension through the introduction of late-time new physics. }
\end{enumerate}
\tk{For all the above, we use the default likelihoods available with} \verb|Cobaya|\kb{, which include full covariance matrices accounting for statistical and systematic errors for Planck 2018 CMB high--$\ell$ (TTTEEE), low-$\ell$ (lowl + lowE) \cite{Aghanim:2018eyx, Planck:2019nip}, and lensing measurements \cite{Planck:2018lbu}, 
BOSS DR 12 \cite{BOSS:2016wmc}, and the Pantheon Supernovae sample \cite{Pan-STARRS1:2017jku}, which can be found in the respective references and their supplemental materials. 
For the BAO distance ratio measurement from  6dFGS at $z_{\rm eff} = 0.106$, we use
$r_s(z_d)/DV (z_{\rm eff}) = 0.336 \pm 0.015$ \cite{Beutler:2011hx}, and for the BAO measurement from the SDSS Main Galaxy
Sample at  $D_V (z_{\rm eff} = 0.15) = (664 \pm 25)(r_d/r_{d,fid})$ we use the full likelihood as provided in Table 3 in \cite{Ross:2014qpa}.}

This combination of data has several important constraining properties. 
CMB and BAO together break geometric degeneracies by measuring the sound horizon at multiple redshifts. 
Their agreement under \LCDM\ also constrains new physics in the redshifts between the CMB and BAO. 
Supernovae from Pantheon strongly constrain new physics over their own redshift range, altogether largely excluding new physics in the late universe. 
Beyond providing these constraints, these datasets are also consistent under a \LCDM\ cosmology, providing a powerful point of comparison for any new physics - new physics should not introduce a tension where previously none existed. 

We term the above combination our \textit{``baseline"} datasets,
\tk{to which we eventually add } 
either or both: 
\begin{enumerate}
    \setcounter{enumi}{3}
    \item the latest SH0ES measurement of the present day Hubble rate $H_0 = 73.04 \pm 1.04 \, \text{km}\text{s}^{-1}\text{Mpc}^{-1}$
    \cite{Riess:2021jrx}, and
    \item the Dark Energy Survey Year 1 (DES Y1) galaxy lensing and clustering measurements \cite{DES:2017myr, DES:2017tss}. 
\end{enumerate}

We include the SH0ES $H_0$ measurement to mitigate prior volume effects \cite{Smith:2020rxx,Herold:2021ksg}, and to evaluate whether our model can reconcile the most discrepant local $H_0$ measurement with the one derived from the CMB. 
We focus on the local late-time measurement resulting in the largest Hubble tension, though we do note that other local measurements with lower $H_0$ exist \cite{Freedman:2019jwv, Riess:2020fzl}. 
\kb{We compare the DA EDE cosmology to $\Lambda$CDM, but we note that \textit{baseline}+$H_0$ is discrepant under $\Lambda$CDM. 
Therefore, the target goodness-of-fit to CMB for this extended model is a \LCDM\ fit to \textit{baseline}. 
}

We include DES Y1 data to study the impact of thermal friction on large-scale structure observations and the LSS tension, wherein \LCDM\ combined with the Planck CMB overpredicts the amplitude of matter fluctuations $\sigma_8$ relative to late-universe measurements relying on weak lensing and galaxy clustering \cite{DES:2017myr, Hildebrandt:2018yau, HSC:2018mrq, KiDS:2020suj, DES:2021wwk}. 
Although DES Y1 data is in $\sim 2 \sigma$ agreement with the Planck \LCDM\ CMB, this dataset nonetheless helps constrain EDE models that generically increase $\omega_{\rm cdm}$ in order to accommodate a larger $H_0$ \cite{Hill:2020osr}.

\tk{Hence, we study DA EDE first with \textit{baseline} alone, and then under optimistic circumstances - with SH0ES and DES forcing a solution to the $H_0$ and $S_8$ tensions.} 
\kb{As we will show in the next section, we find that for every dataset combination 
the best-fit DA EDE solution matches onto a cosmology with extra self-interacting dark relativistic species, a model that requires only one extra parameter rather than the three introduced in the DA EDE cosmology. } 

Recent work has shown that CMB data from the Atacama Cosmology Telescope (ACT) Data Release 4 combined with large-scale Planck TT, Planck CMB lensing, and BAO data prefers the existence of EDE even without the inclusion of local $H_0$ measurements \cite{Hill:2021yec}. 
This is confirmed by independent tests and data from the South Pole Telescope (SPT) \cite{Poulin:2021bjr, Lin:2020jcb, Chudaykin:2020igl, Chudaykin:2020acu}.
This preference is driven by ACT's TE and EE power spectrum data, and the exclusion of the tight constraints arising from Planck's high-$\ell$  TT data, and is supported by data from the South Pole Telescope (SPT) \cite{Hill:2021yec}. 
We do not include these data sets here as it remains unclear whether these differences in support for an EDE cosmology arise from new physics or systematics \cite{Hill:2021yec, Poulin:2021bjr}.

\subsection{Priors}
\label{sec:priors}

Using the data combinations outlined above, we constrain three cosmologies - \LCDM, thermal friction as described in Sec.~\ref{sec:thermal friction} (labeled DA EDE), and varying the massless neutrino degrees of freedom $\Neff$ (labeled $\Neff$). 
For all three, we constrain the standard $\Lambda$CDM parameters:  the baryon density $\Omega_b h^2$, the cold dark matter density $\Omega_ch^2$, the curvature spectrum amplitude $A_s$ at $k = 0.05 \text{Mpc}^{-1}$ and its tilt $n_s$, the reionization optical depth $\tau_\text{reio}$, and the Hubble rate today $H_0$, all with non-informative priors. 

For \LCDM\ and DA EDE, we adopt the standard neutrino description, with one massive neutrino with minimal mass $m_{\nu} = 0.06 \, \text{eV}$ and two massless neutrinos. 
Allowing the number $N_{\rm ur}$ of massless neutrinos to vary as $N_{\rm ur} \in [0, 5]$, the cosmology labeled $\Neff$ adds one additional parameter to \LCDM. 
This cosmology also holds the number of massive neutrinos constant at 1, with a mass of $m_\nu = 0.06$ eV. 

The additional theory parameters beyond \LCDM\ in DA EDE are the constant thermal friction $\Upsilon$, the scalar field mass $m$ both in units of $[$Mpc$]^{-1}$ and the initial value $\phi_i$ of the scalar field in units of $[M_{\text{Pl}}]$. 
For our analysis, we reparameterize the model in terms of effective parameters ($m$, $\frac{m^2}{\Upsilon}$, $f_{\ede}$),
where $f_{\ede}$ denotes the maximal fractional amount of scalar field and dark radiation energy density combined, 
and $\frac{m^2}{\Upsilon}$ maps onto the often used phenomenological parameter $z_c$, the redshift at which $f_{\ede}$ peaks \kb{as defined in Eq. \ref{eq:fede}}. 

\begin{figure}
    \centering
    \includegraphics[width=0.49\textwidth]{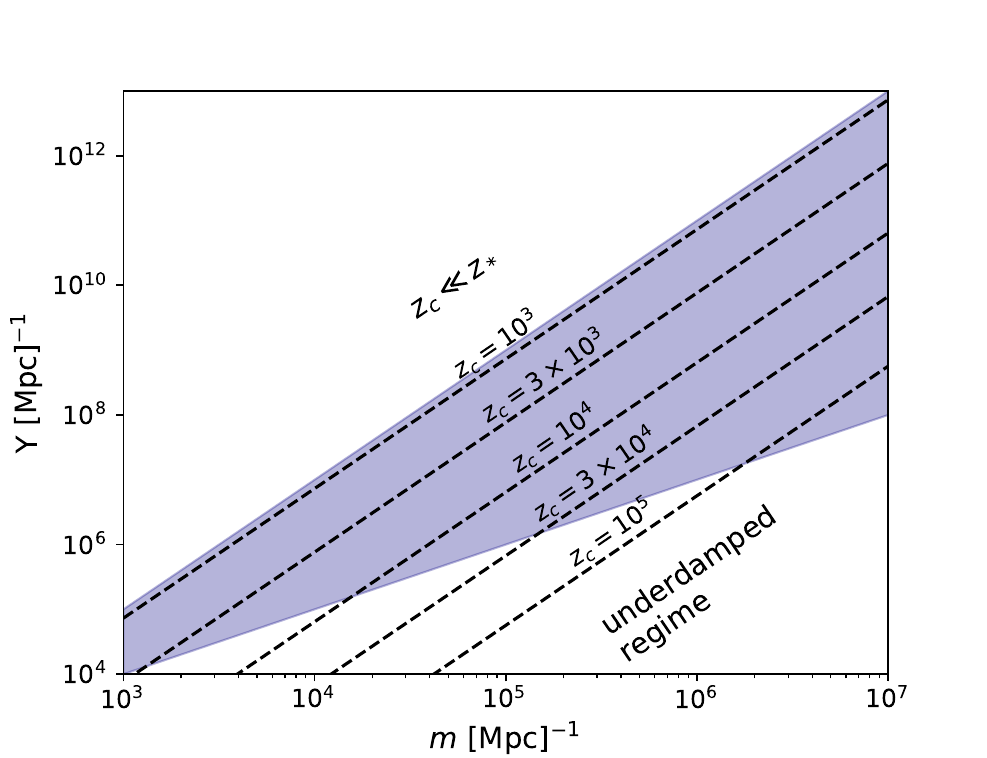}
    \caption{
         The shaded region shows the allowed parameter space for $\Upsilon$ and $m$, given the priors specified in Table~\ref{tab:priors}. 
         The dashed curves correspond to constant $z_c$ as labeled. 
         This region probes a range from $770 < z_c < 2.2 \times 10^5$.
     }
    \label{fig:prior_plot}
\end{figure}

\begin{table}[]
\centering
\begin{tabular}{|l| c |} 
	\hline 
	Parameter   & Prior      \\  
	\hline 
	\hline
	$\log_{10} m $  &[3, 7]         \\
	$ \log_{10} \frac{m^2}{\Upsilon}$  & [1, $\log_{10} m - 1$]           \\
	$f_{\text{ede}}$      &[0.001 ,0.2]    \\
	\hline 
\end{tabular} 
\caption{ 
        Priors for thermal friction parameters, where $m$ and $m^2/\Upsilon$ have units of [Mpc]$^{-1}$. 
        These priors allow for the following ranges in derived parameters: $770 < z_c < 2.2 \times 10^5$ and 
        $10^{-5} < \frac{\phi_i}{M_{\text{Pl}}} < 10^{-2} $.
        }
\label{tab:priors}
\end{table}

We impose the priors in Table~\ref{tab:priors}, where the prior on $m^2/\Upsilon$ has the additional prior of $m \ll \Upsilon$ to remain in the overdamped regime. 
The purpose of this is two-fold.  
Primarily, this forces the scalar field to instantaneously dump its energy density into the dark radiation, which then dilutes away as $a^{-4}$. 
Early dark energy scenarios only alleviate the Hubble tension if their energy density dilution is at least as fast as that of radiation \cite{Poulin:2018dzj}, a box ticked by this additional prior. 
Another valuable advantage of this choice is that our results become broadly applicable to all other choices of potentials, as the evolution of the scalar is dominated by thermal friction, making the specifics of its potential irrelevant. 
Hence the fine-tuning of EDEs is guaranteed to be alleviated by the construct of overdamped DA EDE itself - regardless of potential.  

In Fig. \ref{fig:prior_plot}, we show how the prior on our effective parameters translates to allowed ranges on the theory parameters $m$ and $\Upsilon$. 
The curves of constant $z_c$ show the redshifts that this permitted region probes. 
Further expanding our prior range in $m$ to include larger values increases the maximum possible $z_c$. 
However, we limit ourselves to $z_c \lesssim 2\times10^5$, as data loses sensitivity to higher injection redshifts \cite{Linder:2010wp,Karwal:2016vyq}. 
We find that our model asymptotes to these high injection redshifts, approaching the extra-radiation regime, and explore this further in Sec.~\ref{sec:results_no_EDE_solution}. 

\section{Results}
\label{sec:results}

For each combination of data sets, we compare the three cosmologies \LCDM, DA EDE and $\Neff$.
As mentioned before, $\Neff$ provides a 1-parameter approach to adding extra free-streaming radiation to a \LCDM\ universe. 
We seek to juxtapose our model with $\Neff$ to compare the benefits or drawbacks of adding a more complicated dark radiation which couples to an EDE-like scenario. 
We illustrate the Hubble and LSS tensions with the parameters $H_0$ and $S_8 = \sigma_8 \sqrt{\frac{\Omega_m}{0.3}}$ respectively. 
Our objectives are to quantify:

\begin{enumerate}[label=(\roman*)]
    \item how the addition of thermal friction impacts the Hubble tension, 
    \item the goodness of fit of thermal friction in comparison to our reference cosmologies, and 
    \item the impact of thermal friction on the LSS tension.
\end{enumerate}

Since we find our best fit to prefer asymptotically early injection of dark radiation, we further investigate 
\begin{enumerate}[label=(\roman*)]
    \setcounter{enumi}{3}
    \item at what redshift $z_c$ the data looses sensitivity to DA EDE injection time, and 
    \item why the EDE-like regime of thermal friction is not preferred by data.
\end{enumerate}
We discuss (i) in \kb{Sec.~\ref{sec:results_baseline} and} Sec.~\ref{sec:results_H0}, (iii) in Sec.~\ref{sec:results_LLS}, (iv) in Sec.~\ref{sec:results_higher_z_c} and (v) in Sec.~\ref{sec:results_no_EDE_solution}. 
We explore (ii) throughout.

\subsection{Data consistent under \LCDM}
\label{sec:results_baseline}

\tk{
Planck CMB data, Pantheon supernova data and the BAO measurements outlined in Sec.~\ref{sec:datasets} are concordant under a \LCDM\ cosmology. 
As \LCDM\ is a nested model in both DA EDE and varying $\Neff$, we do not expect to introduce new tensions between these data. 
Ideally, the new physics introduced to resolve the $H_0$ and $S_8$ cosmic tensions would succeed without the inclusion of SH0ES or weak-lensing data. 
While the inability of $\Neff$ to achieve this is documented \cite{Planck:2018vyg, Poulin:2018zxs}, here we test if DA EDE fit to these concordant data alone predicts a higher $H_0$ or lower $S_8$. 
}

\begin{figure}
    \centering
    \includegraphics[width=0.49\textwidth]{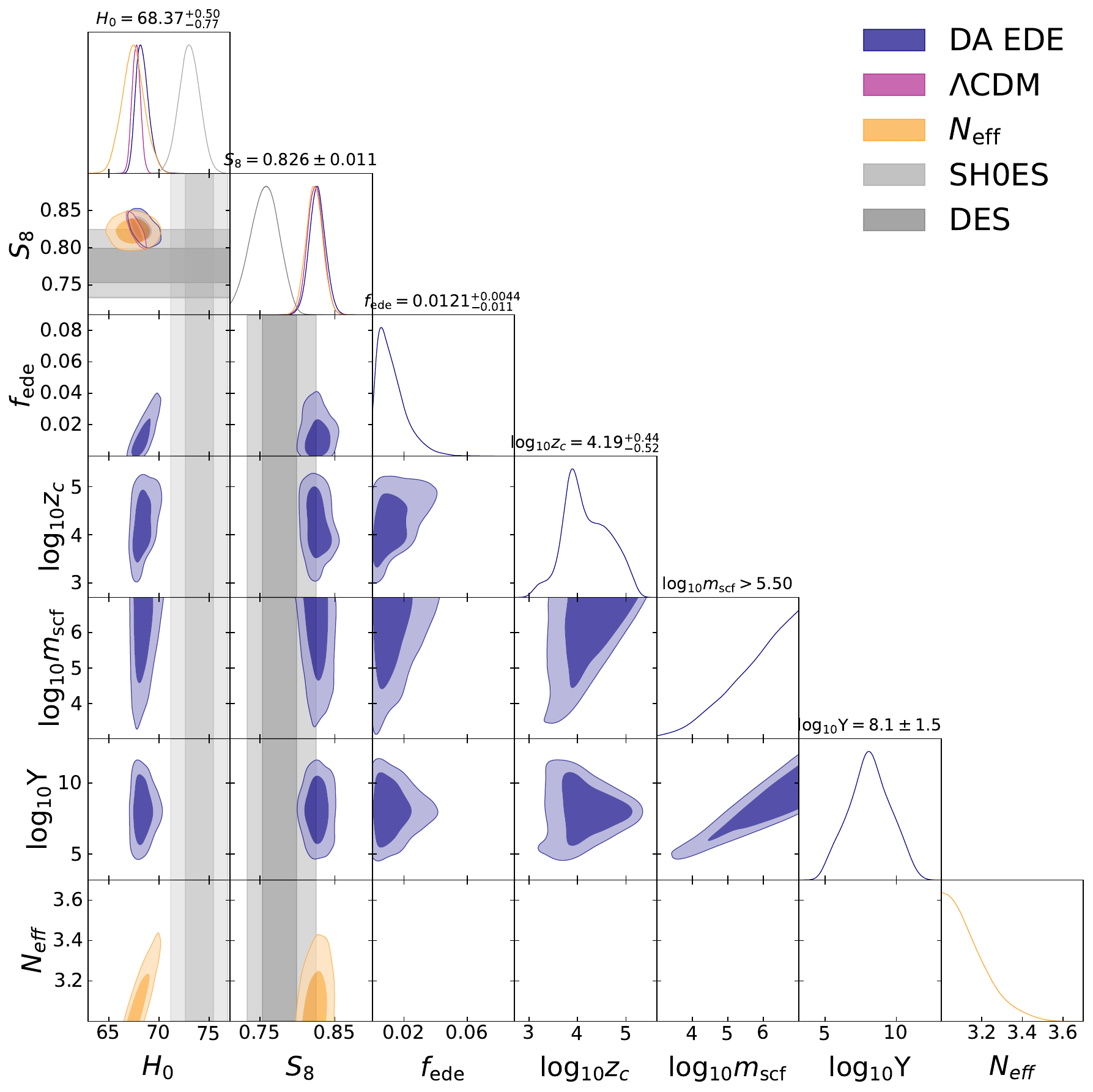}
    \caption{
    \tk{
    MCMC posteriors for \LCDM, $N_{\rm eff}$ and DA EDE, fitting to our \textit{baseline} data combination Planck, BAO and SNe, which are concordant under a \LCDM\ cosmology. 
    In gray bands, we show the SH0ES constraint on $H_0$ and the DES Y1 constraint on $S_8$. 
    The labels at the top of each column are the 1D marginalized posteriors for each parameter in a DA EDE cosmology. 
    }
    }
    \label{fig:base}
\end{figure}

\tk{
2D marginalized posteriors for the DA EDE parameters, $\Neff$ and the two parameters $H_0$ and $S_8$ that quantify the tensions are shown in Fig.~\ref{fig:base}, while Table~\ref{tab:baseline_small_results} records the 1D marginalized posteriors and best-fits values of the tension parameters. 
The Hubble tension becomes apparent between the 2$\sigma$ SH0ES band shown in light gray, and the \LCDM\ contour. 
The smaller LSS tension is represented here by DES Y1 constraints on $S_8$, in darker gray. 
While DA EDE predicts a higher $H_0$ than \LCDM, the increase is small, and the model does not provide a solution to either tension. 
}

\begin{table}[h]
    \centering
    \begin{tabular}{|l|c|c|}
    \hline
    Model   &   $H_0$ [km/s/Mpc]  & $S_8$ \\
    \hline
    \hline
    \LCDM  &    $67.73(67.86)\pm 0.42$ 
            &   $0.8239(0.8226) \pm 0.0104$ 
            \\
    DA EDE  &   $68.37(68.14)^{+0.50}_{-0.77}$ 
            &   $0.8265(0.8308) \pm 0.0107$ 
            \\
    $\Neff$ &   $67.4(67.3)\pm 1.1$ 
            &   $0.8224(0.8237) \pm 0.0109$ 
            \\
    \hline
    \end{tabular}
    \caption{
    \tk{
    1D marginalized posteriors of measurements quantifying the two cosmological tensions, showing the mean (bestfit) $\pm 1 \sigma$, fitting to \textit{baseline} data. 
    }
    }
    \label{tab:baseline_small_results}
\end{table}

\tk{
Nonetheless, the 2D contours and 1D marginalized likelihoods exhibit an interesting feature - while varying $\Neff$ only increases the error on $H_0$ without shifting its central value, DA EDE both shifts the central value and broadens the error. 
Moreover, DA EDE only increases the predicted $H_0$ relative to \LCDM, while $\Neff$ broadens the $H_0$ contours to both higher and lower values. 
However, this feature arises from our choice of priors for DA EDE and not the model itself. 
Pre-recombination EDE ($z_c > z_*$) can alleviate the Hubble tension, but post-recombination EDE ($z_c < z_*$) worsens it \cite{Poulin:2018dzj}. 
As our primary interest here in DA EDE is in its potential to resolve tensions, we limit $z_c$ to the range of interest, leading to contours that only increase $H_0$ realtive to \LCDM. 
}

\tk{
Choices of priors also dominate posteriors on the $\Upsilon-m$ plane. 
The preferred region for $f_{\rm ede} < 5\%$ at $2\sigma$ with the mean at $f_{\rm ede} \simeq 1\%$. 
For such small amounts of DA EDE, data become insensitive to its properties and $\Upsilon$ and $m$ become unconstrained, their contours effectively tracing the prior in Fig.~\ref{fig:prior_plot}. 
}

\tk{
Finally, Table~\ref{tab:baseline_chis} shows the CMB and total $\chi^2$ for all three cosmologies fit to \textit{baseline}. 
We also calculate the Bayesian information criteria (BIC) \cite{Trotta:2008qt} for all models and present $\Delta$BIC relative to \LCDM\ fit to the same data, with the model with minimum BIC being preferred (see App.~\ref{app:BIC} for details).  
While generally one expects a $\chi^2$ improvement equal to the number of additional parameters, this is difficult to quantify for non-linear models \cite{Andrae:2010gh}. 
Specifically in DA EDE, prior-volume effects change the effective number of additional parameters across the parameter space. 
As $f_{\rm ede} \rightarrow 0$, the model adds one extra parameter to \LCDM, but for larger $f_{\rm ede}$, data gains sensitivity to the existence of DA EDE and the number of additional parameters can rise to 3. 
Analysis with additional data in the next subsections shows a preference for the regime where DA EDE is injected at very early redshifts $z_c > 10^4$, asymptoting to an extra self-interacting dark radiation solution \cite{Blinov:2020hmc}. 
This regime also reduces the effective number of extra parameters to 1 - the amount of extra dark radiation. 
Posteriors for DA EDE fit to \textit{baseline} prefer the $f_{\rm ede} \rightarrow 0$ regime, such that DA EDE effectively adds one extra parameter to \LCDM, same as $\Neff$. 
While for DA EDE, the improvement is close to that expected $\Delta \chi^2_{\rm total} \simeq -1$, $\Neff$ finds essentially no improvement. 
This is consistent with analyses that allow $\Neff$ to vary, which find it consistent with the \LCDM\ value $\Neff = 3.046$ \cite{Planck:2018vyg}. 
Lastly, the BIC shows preference for \LCDM\ over both extensions. 
}

\begin{table}[h]
    \centering
    \begin{tabular}{|l|c|r|c|c|}
    \hline
    Model   &   $\chi^2_{\rm CMB}$  & $\chi^2_{\rm total}$ & \kb{$\Delta$BIC}\\
    \hline
    \hline
    \LCDM   &   2773.2 
            &   3813.54 
            &   \tk{0}
            \\
    DA EDE  &   2772.2
            &   3812.82 
            &   \tk{23.68}
            \\
    $\Neff$ &   2772.7
            &   3813.7
            &   \tk{8.29}
            \\
    \hline
    \end{tabular}
    \caption{\tk{The goodness of fit when fitting to the \textit{baseline} data combination, and the $\Delta$BIC relative to BIC$_{\Lambda {\rm CDM}} = 4033.13$.
    } 
    }
    \label{tab:baseline_chis}
\end{table}

\subsection{The Hubble tension}
\label{sec:results_H0}

\begin{figure}
    \centering
    \includegraphics[width=0.49\textwidth]{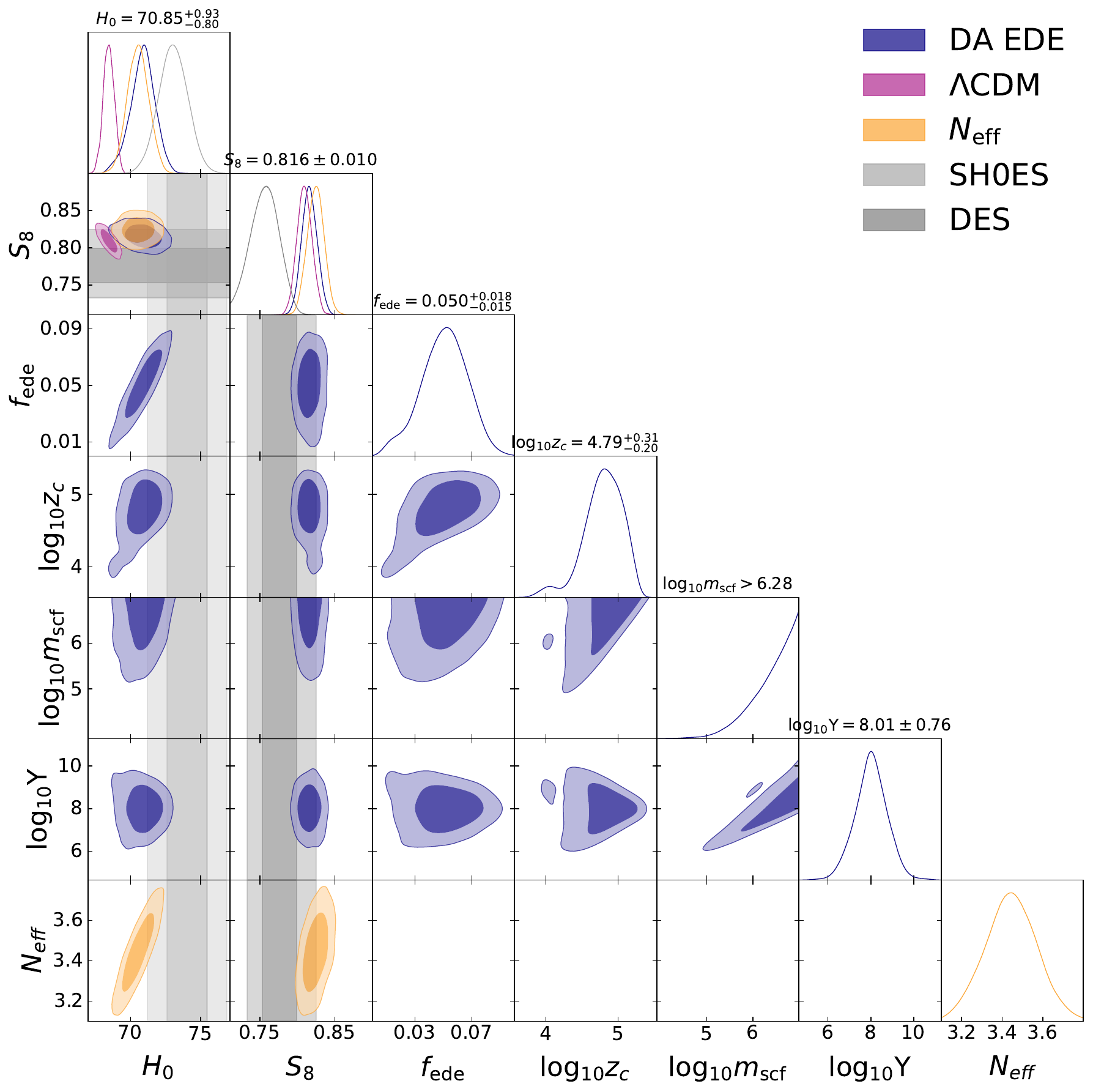}
    \caption{
    MCMC posteriors for \LCDM, $N_{\rm eff}$ and DA EDE, while fitting to \textit{baseline} + SH0ES, following the same conventions as Fig.~\ref{fig:base}.  
    \tk{Note that under \LCDM\, the data combined here are discrepant, but we show \LCDM\ contours to draw a comparison with $\Neff$ and DA EDE run on the same data combination. }
    }
    \label{fig:base_H0}
\end{figure}

In Fig.~\ref{fig:base_H0}, we show the 2D posteriors obtained by fitting to our \textit{baseline} data sets plus the local Hubble measurement \cite{Riess:2021jrx} (\textit{baseline}+$H_0$), following the methodology described in Sec.~\ref{sec:datasets}. 
DA EDE (purple) can be seen to reduce the Hubble tension while leaving $S_8$ roughly unchanged. 
Comparing to $\Neff$, DA EDE finds a larger $H_0$ \tk{by $1.18$km/s/Mpc} and smaller $S_8$ \tk{by $0.0071$}, gaining improvements in both tensions over simply varying the number of massless neutrinos, as also shown in Table~\ref{tab:baseline_h_small_results}. 
Data also show preference for $f_\ede > 0$ at $>2\sigma$, with a best fit of $f_\ede = 6.3\%$. 
This best fit is similar to other EDEs that dilute like radiation as $a^{-4}$ \cite{Poulin:2018cxd, Agrawal:2019lmo}. 

\begin{table}[h]
    \centering
    \begin{tabular}{|l|c|c|}
    \hline
    Model   &   $H_0$ [km/s/Mpc]  & $S_8$ \\
    \hline
    \hline
    \LCDM  &    $68.44(68.53)\pm 0.39$   
            &   $0.8093(0.8095) \pm 0.0100$ 
            \\
    DA EDE  &   $70.85(71.43)^{+0.93}_{-0.80}$  
            &   $0.8159(0.8157) \pm 0.0102$ 
            \\
    $\Neff$ &   $70.53(70.25)\pm 0.76$ 
            &   $0.8241(0.8228) \pm 0.0111$
            \\
    \hline
    \end{tabular}
    \caption{1D marginalized posteriors of measurements quantifying the two cosmological tensions, showing the mean (bestfit) $\pm 1 \sigma$, fitting to \textit{baseline}+$H_0$. 
    }
    \label{tab:baseline_h_small_results}
\end{table}

Projecting the amount of extra radiation energy density of our best fit of $f_{\text{ede}} =6.3\%$ onto the amount of energy density that would be provided by extra neutrino species, we find a corresponding $\Delta \Neff =0.47$. 
In this mapping, we set 
\begin{equation}
    \rho_{\text{r}} = \rho_{\gamma}\left(1 + \frac{7}{8} \Neff \left(\frac{4}{11}\right)^{\frac{4}{3}}  \right) \,,
\end{equation}
with $\Neff = 3$, and then solve for $\Delta \Neff$ such that 
\begin{equation}
    \Delta \Neff = \frac{8}{7}\left(\frac{11}{4}\right)^{\frac{4}{3}}  \left(1 + \frac{7}{8} \Neff \left(\frac{4}{11}\right)^{\frac{4}{3}}  \right)f_{\text{ede}} \,.
\end{equation}
Of course, this is a crude mapping, with the comprehensive differences between the $\Neff$ and DA EDE cosmologies fully quantified by the posteriors shown in Fig. \ref{fig:base_H0} and \ref{tab:full_base_h}. 

Another interesting feature of these posteriors is that the preferred EDE injection time is at high $z_c > 10^4$, with our best-fit $z_c = 9.1 \times 10^4$.
The shape of the $z_c$ contours in Fig. \ref{fig:base_H0} corresponds to the possible redshifts that can be explored subject to satisfying the priors on $\Upsilon$.
Typically, EDE modifies the expansion history close to matter-radiation equality, localized in redshift around $z_{\rm eq} \simeq 3300$, maximizing its impact on the sound horizon and hence the predicted $H_0$. 
However, in DA EDE, we find that the preferred region in $z_c$ occurs at much higher redshifts, with data effectively favoring the additional radiation energy density over an EDE-like injection. 
This corresponds to a constant fractional increase in energy density throughout radiation domination, which fades as matter takes over \cite{Knox:2019rjx}. 
Indeed, our results show a preference for an asymptotic extra-radiation solution, in which self-interacting dark radiation has always been part of cosmic history, throughout the redshifts (up to $z \sim  10^5$) that data is sensitive to \cite{Karwal:2016vyq, Linder:2010wp}.
The ability of this asymptotic solution to ease the Hubble tension without disturbing, \kb{but also without further solving} $S_8$ is in agreement with other investigations of the effects of self-interacting 
radiation on the Hubble tension \cite{Kreisch:2019yzn, Park:2019ibn,Blinov:2020hmc, Brinckmann:2020bcn}. 

Finally, we consider the impact to the $\chi^2$ for the CMB and SH0ES in Table~\ref{tab:baseline_h_chis}. 
Driven by the inclusion of the local SH0ES measurement, DA EDE and $\Neff$ both predict a higher $H_0$. 
Studies have shown that this increase in $H_0$ is not compensated for by $\Neff$ in a manner that either decreases or maintains $\chi^2_{\rm CMB}$ \cite{Aghanim:2018eyx, Riess:2016jrr, Poulin:2018zxs}, as generally achieved by EDEs.
\tk{From the total $\chi^2$'s in Table~\ref{tab:baseline_h_chis}, DA EDE provides an improvement of $\Delta \chi^2 \simeq 13$ for only three additional degrees of freedom, relative to the expectation of $\Delta \chi^2 = 1$ per additional parameter. 
Although naively this would imply that the model performs well, this improvement comes entirely from a higher $H_0$, and indeed at cost to the CMB $\chi^2$. 
}

\begin{table}[h]
    \centering
    \begin{tabular}{|l|c|r|c|c|}
    \hline
    Model   &   $\chi^2_{\rm CMB}$  & $\chi^2_{H_0}$ & $\chi^2_{\rm total}$ & \kb{$\Delta$BIC}\\
    \hline
    \hline
    \LCDM  &   2777.5   
            &   18.8
            &   3836.56
            &   \kb{0}
            \\
    DA EDE  &   2780.3 
            &   2.4
            &   3823.55
            &   \kb{11.39}
            \\
    $\Neff$ &   2780.0
            &   7.2
            &   3827.70
            &   \kb{-0.73}
            \\
    \hline
    \end{tabular}
    \caption{The goodness of fit to CMB data and SH0ES, while cumulatively fitting to \textit{baseline}+$H_0$. For reference, \LCDM\ fit just to \textit{baseline} has $\chi^2_{\rm CMB} = 2772.6$. 
    \tk{
    The $\Delta$BIC are shown relative to BIC$_{\Lambda {\rm CDM}} = 4056.16$.
    }
    }
    \label{tab:baseline_h_chis}
\end{table}

Note that the \LCDM\ $\chi^2_{\rm CMB}$ reported in Table~\ref{tab:baseline_h_chis} is already greater than a \LCDM\ fit to just the concordant \textit{baseline} datasets. 
\tk{
Under \LCDM, \textit{baseline} data sets are consistent with each other, but incompatible with SH0ES. 
A \LCDM\ fit to \textit{baseline}+SH0ES worsens the fit to the CMB by $\Delta \chi^2 \simeq 5$. 
The attraction of EDEs stems from their ability to predict a high $H_0$ compatible with SH0ES, while providing a fit to the CMB competitive with \LCDM\ fit only to concordant data sets. 
We do not find these important qualities in DA EDE, which reduces the Hubble tension at the cost of CMB $\chi^2$. 
} 

\tk{
Moreover, the BIC shows preference for \LCDM\ over DA EDE despite its improvement in total $\chi^2$ in Table~\ref{tab:baseline_h_chis}, while $\Neff$ has a BIC comparable to \LCDM. 
Ultimately, we find similarities with the weaknesses of the $\Neff$ solution, in that, the CMB $\chi^2$ is worsened at the cost of improving the Hubble tension. 
}

\subsection{Impact of LSS measurements}
\label{sec:results_LLS}

Searching for a new concordance model of cosmology to replace \LCDM, we not only need to address the Hubble tension, but also resolve the LSS tension. 
In this section, we consider the impact of including DES Y1 data in our analysis on cosmological parameter constraints as well as goodness-of-fits. 

\begin{figure}
    \centering
    \includegraphics[width=0.49\textwidth]{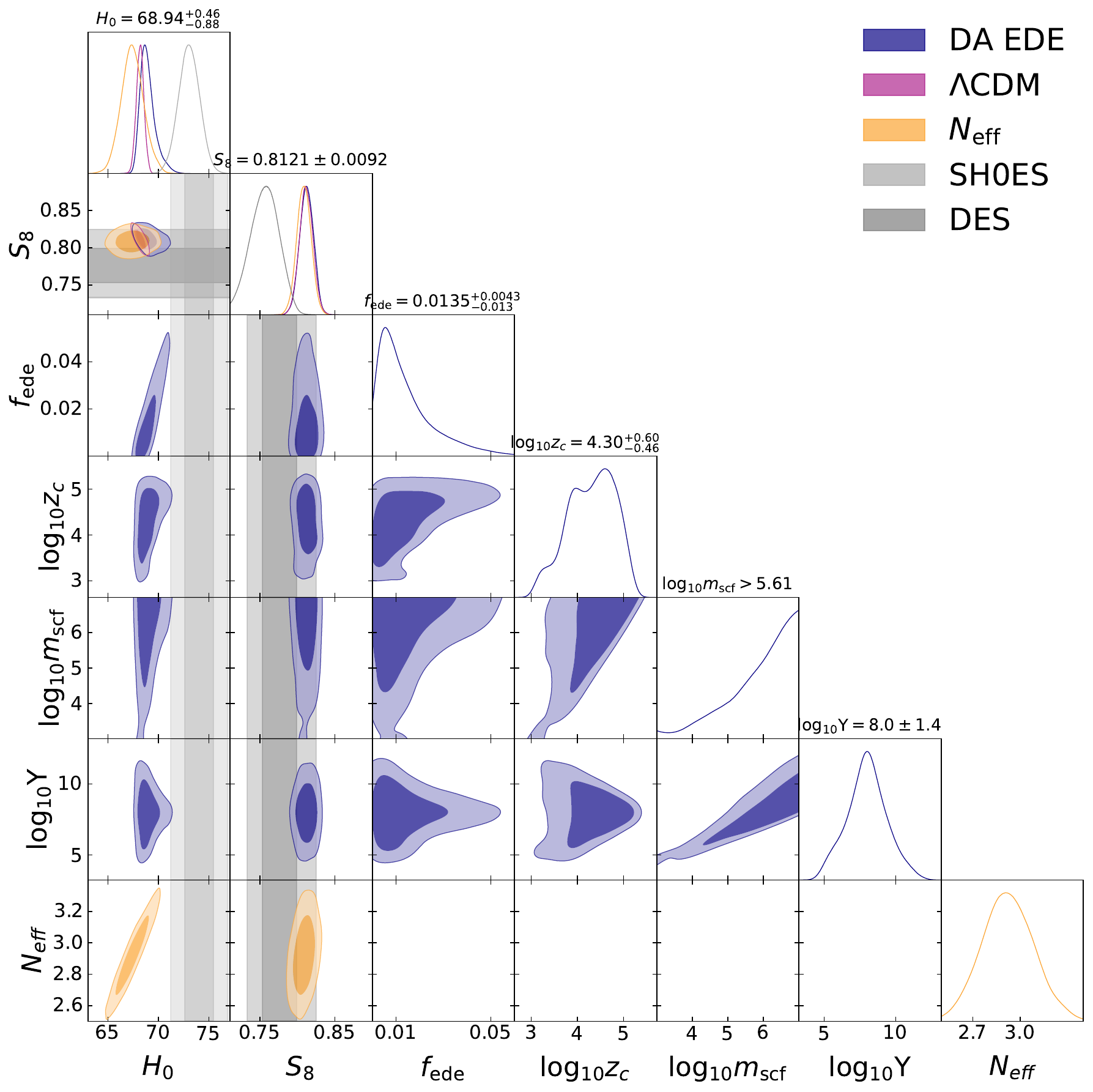}
    \caption{
    Posteriors for \LCDM, \LCDM\ with variable $N_{\rm eff}$ and DA EDE, while fitting to \textit{baseline} and DES Y1. 
    As in Fig.~\ref{fig:base_H0}, we show the SH0ES constraint on $H_0$ and the DES Y1 constraint on $S_8$ and 1D marginalized posterior for each parameter in a DA EDE cosmology at the top of each column.  }
    \label{fig:base_des}
\end{figure}

We begin by adding just DES to our \textit{baseline} data, to determine if LSS data suppresses the amount of DA EDE as it does for various other EDE models \cite{Ivanov:2020ril, Hill:2020osr} (but see \cite{Smith:2020rxx} for an alternate perspective). 
These results are shown in Fig.~\ref{fig:base_des} and Table~\ref{tab:base_des_small_results}. 

\begin{table}[h]
    \centering
    \begin{tabular}{|l|c|c|}
    \hline
    Model   &   $H_0$ [km/s/Mpc]  & $S_8$ \\
    \hline
    \hline
    \LCDM   &   $68.19(68.08)\pm 0.38$ 
            &   $0.8115(0.8128) \pm 0.0091$ 
            \\
    DA EDE  &   $68.94(68.38)^{+0.46}_{-0.88}$ 
            &   $0.8120(0.8133) \pm 0.0091$ 
            \\
    $\Neff$ &   $67.4(67.2)\pm 1.1$ 
            &   $0.8086(0.8078) \pm 0.0094$ 
            \\
    \hline
    \end{tabular}
    \caption{1D marginalized posteriors of measurements quantifying the two cosmological tensions, showing the mean (bestfit) $\pm 1 \sigma$, fitting to \textit{baseline}+DES Y1.}
    \label{tab:base_des_small_results}
\end{table}

The addition of DES data without the SH0ES $H_0$ prior does not by itself prefer non-zero $f_\ede$, but nor does DES data exert a strong enough pull to significantly shift the predicted $S_8$ from \textit{baseline}+$H_0$.
This is unlike including the SH0ES $H_0$ prior, which substantially shifts the predicted $H_0$ relative to \textit{baseline}+DES. 
This is unsurprising, given the smaller tension between DES and \textit{baseline}. 
Ultimately, we find constraints for both DA EDE and $\Neff$ that include \LCDM, but with broader error bars. 

\begin{table}[h]
    \centering
    \begin{tabular}{|l|c|c|c|c|}
    \hline
    Model   &   $\chi^2_{\rm CMB}$  & $\chi^2_{\rm DES}$ & $\chi^2_{\rm total}$ & \kb{$\Delta$BIC}\\
    \hline
    \hline
    \LCDM  &   2774.1   
            &   509.3
            &   4323.45
            &   \kb{0}
            \\
    DA EDE  &   2774.7
            &   509.4 
            &   4324.12
            &   \kb{25.1}
            \\
    $\Neff$ &   2776.0
            &   508.1
            &   4324.17
            &   \kb{8.86} 
            \\
    \hline
    \end{tabular}
    \caption{The goodness of fit to CMB and DES data, while cumulatively fitting to \textit{baseline}+DES. 
    \tk{
    We show $\Delta$BIC relative to BIC$_{\Lambda {\rm CDM}} = 4706.11$ .
    }
    }
    \label{tab:baseline_des_chis}
\end{table}

\begin{figure}
    \centering
    \includegraphics[width=0.49\textwidth]{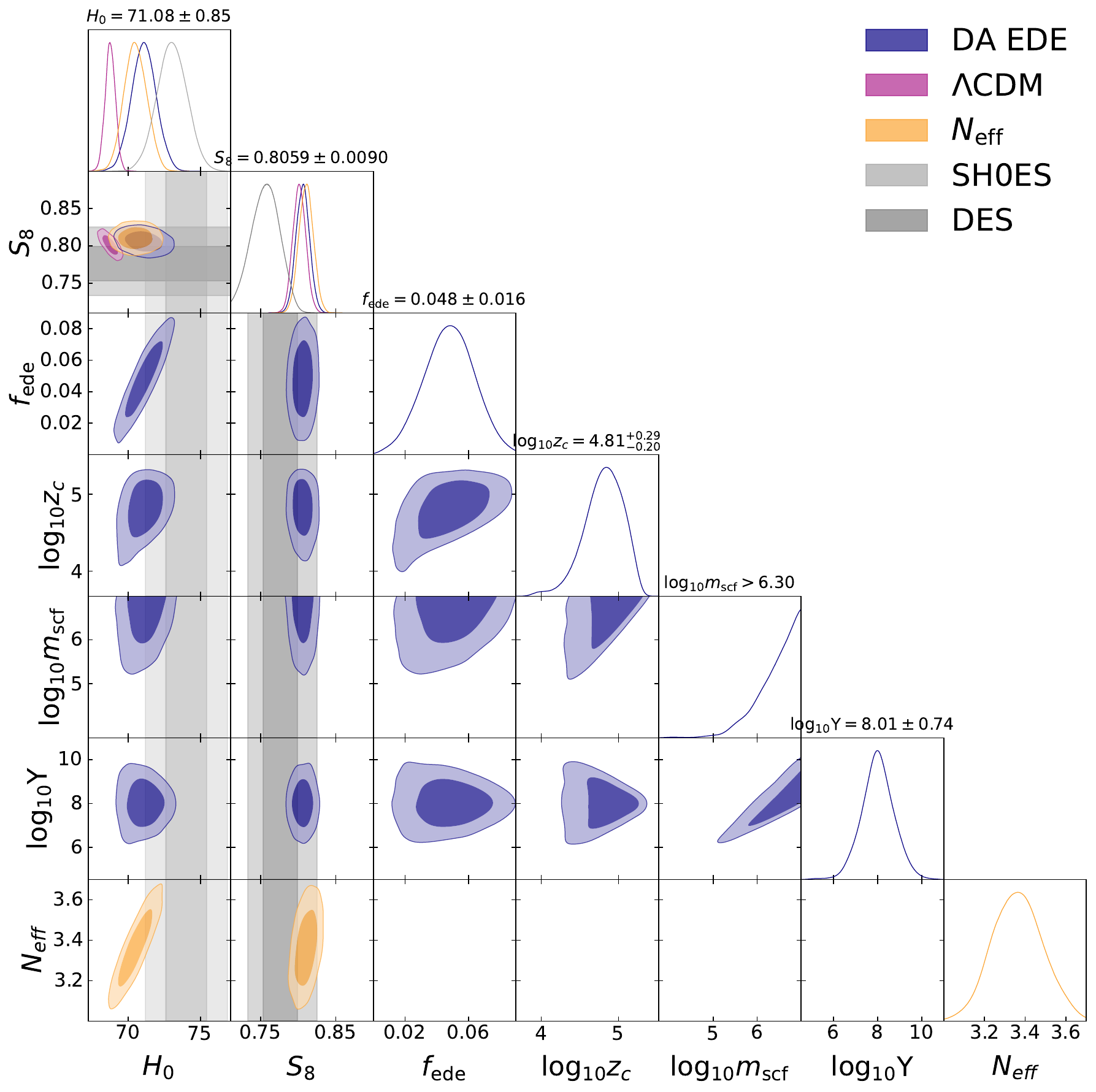}
    \caption{
    Following the same conventions as Figs.~\ref{fig:base_H0} and \ref{fig:base_des}, we show posteriors for \LCDM, \LCDM\ with variable $N_{\rm eff}$ and DA EDE cosmologies, while fitting \textit{baseline} plus SH0ES and DES Y1. 
    \tk{Note again that under \LCDM\, the data combined here are discrepant. However, the \LCDM contours allow for a direct comparison with the extended cosmologies. }
    }
    \label{fig:base_h_des}
\end{figure}

\begin{figure}
    \centering
    \includegraphics[width=0.49\textwidth]{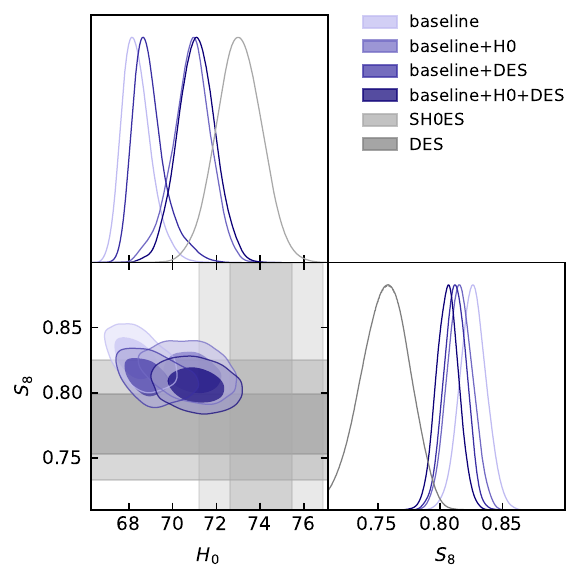}
    \caption{
    Here we show the impact of inclusion of LSS data on DA EDE posteriors, along with the local estimations of $H_0$ by SH0ES and $S_8$ from DES Y1.
    As expected, both tensions are improved most when fitting to both DES and SH0ES simultaneously. }
    \label{fig:all_edes}
\end{figure}

The more interesting posteriors result when we include both DES and SH0ES with our \textit{baseline} data, shown in Fig.~\ref{fig:base_h_des} and Table~\ref{tab:base_h_des_small_results}.
As with \textit{baseline}$+H_0$, we find $>2\sigma$ preference for $f_\ede > 0$, with a smaller best fit of $f_\ede = 4.7\%$ and asymptoting $z_c$ with a best fit of $z_c = 8.1 \times 10^4$. 
But in this case, DA EDE and \LCDM\ both surprisingly find \tk{posteriors with} a higher $H_0$ and a lower $S_8$ than in previous data set combinations, as also shown by Fig.~\ref{fig:all_edes}. 
\tk{
Based on the $\chi^2$'s per individual data set in Tables \ref{tab:baseline_h_chis}, \ref{tab:baseline_des_chis} and \ref{tab:base_h_des_chis}, we explain this as follows.
Alone, the DES data does not exert a strong pull on posteriors as seen from Fig.~\ref{fig:base_des}. 
However, combined with the SH0ES likelihood, both DES and SH0ES exert a stronger pull on posteriors in a direction that optimizes both their $\chi^2$'s, leading to the results in Table~\ref{tab:base_h_des_small_results}. 
}

\begin{table}[h]
    \centering
    \begin{tabular}{|l|c|c|}
    \hline
    Model   &   $H_0$ [km/s/Mpc]  & $S_8$ \\
    \hline
    \hline
    \LCDM   &   $68.76(68.63)\pm 0.36$ 
            &   $0.8013(0.8055) \pm 0.0087$ 
            \\
    DA EDE  &   $71.08(71.06)\pm 0.85$ 
            &   $0.8058(0.8075) \pm 0.0089$ 
            \\
    $\Neff$ &   $70.50(70.86)\pm 0.78$ 
            &   $0.8102(0.8106) \pm 0.0096$ 
            \\
    \hline
    \end{tabular}
    \caption{1D marginalized posteriors of measurements quantifying the two cosmological tensions, showing the mean (bestfit) $\pm 1 \sigma$, fitting to \textit{baseline}+$H_0$+DES. }
    \label{tab:base_h_des_small_results}
\end{table}

In both posteriors that include DES, \LCDM\ and DA EDE $\chi^2$s are comparable, but $\Neff$ worsens the fit to the CMB, as shown in Tables~\ref{tab:baseline_des_chis}-\ref{tab:base_h_des_chis}, with all three models producing worse $\chi^2_{\rm CMB}$ than a \LCDM\ fit to just \textit{baseline} which has $\chi^2_{\rm CMB} = 2772.6$. 
When considering all data sets, DA EDE improves $\chi^2_{H_0}$ over $\Neff$ with a total $\chi^2$ improvement over \LCDM\ of $\Delta \chi^2_{\rm total} = -12.32$ and $\Neff$ of $-6.15$. 
\tk{
While this improvement in goodness-of-fit is greater than the number of new parameters added, 3 and 1 respectively, the BIC disfavors both models over \LCDM\ as seen from Table~\ref{tab:base_h_des_chis}.  
Moreover, 
}
\kb{our best fit corresponds to the regime where DA EDE matches onto the known extra dark radiation solution \cite{Blinov:2020hmc} (with no shear perturbations) which can be fully quantified by just one additional parameter. } 

\kb{The best-fit $H_0$ we obtain for \textit{baseline}+H0 ($H_0=71.43\, \text{km/s/Mpc}$) is slightly larger than that for \textit{baseline}+DES+$H_0$ ($H_0=71.06\, \text{km/s/Mpc}$) and significantly larger than \text{baseline}+DES ($H_0=68.38\, \text{km/s/Mpc}$), coinciding with the amount of $f_{\text{ede}}$. 
The mean value for $H_0$ is largest for \textit{baseline}+$H_0$+DES, but the shift from \textit{baseline}$+H_0$ is not statistically significant. 
}

\begin{table}[h]
    \centering
    \begin{tabular}{|l|c|r|c|c|c|}
    \hline
    Model   &   $\chi^2_{\rm CMB}$  
            &   $\chi^2_{H_0}$ 
            &   $\chi^2_{\rm DES}$
            &   $\chi^2_{\rm total}$
            &   \kb{$\Delta$BIC}
            \\
    \hline
    \hline
    \LCDM   &   2778.4 
            &   18.0
            &   508.0
            &   4344.84 
            &   \kb{0} 
            \\
    DA EDE  &   2778.7
            &   3.6
            &   508.3 
            &   4332.52
            &   \kb{12.11} 
            \\
    $\Neff$ &   2783.3
            &   4.4
            &   508.8
            &   4338.67
            &   \kb{1.97} 
            \\
    \hline
    \end{tabular}
    \caption{The goodness of fit to CMB and DES data, while cumulatively fitting to \textit{baseline}+$H_0$+DES.
    \tk{
    We show $\Delta$BIC relative to BIC$_{\Lambda {\rm CDM}} = 4727.52$ .
    }
    }
    \label{tab:base_h_des_chis}
\end{table}

\subsection{Data are not sensitive to higher $z_c$}
\label{sec:results_higher_z_c}

\begin{figure}
    \centering
    \includegraphics[width=0.49\textwidth]{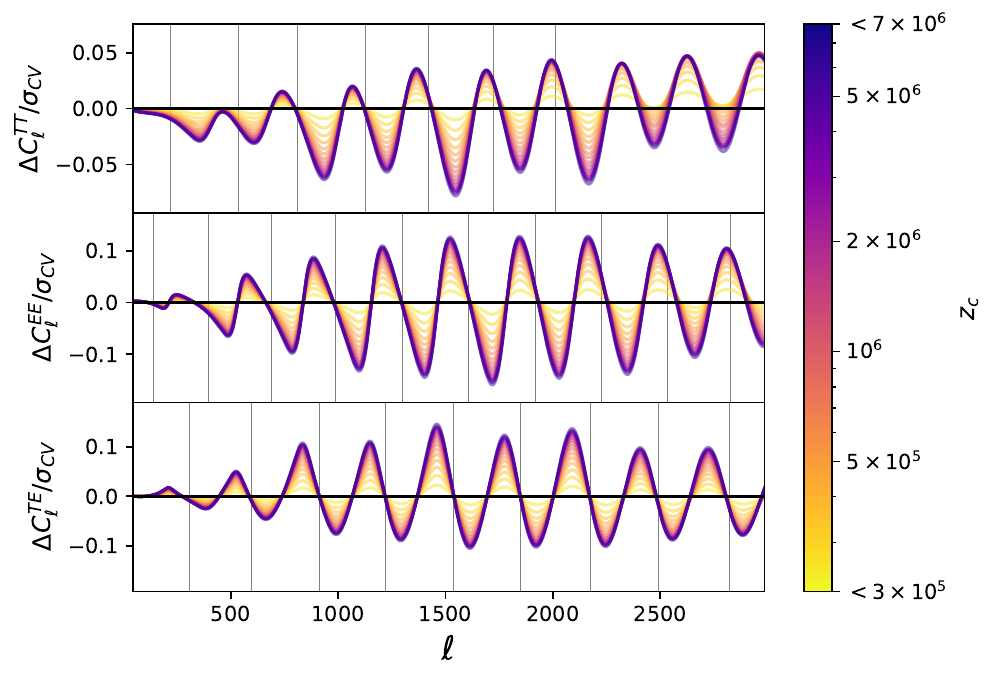}
    \caption{
    We show the CMB residuals in units of cosmic variance for various DA EDE cosmologies. 
    In this format, for features to be observable, they must be different by $\Delta C_\ell ^{XX}/\sigma_{\rm CV} > 1$. 
    These curves fix $m = 10^9$ Mpc$^{-1}$ and all other cosmological parameters at the \textit{baseline}+$H_0$ best fit, varying just $\log (m^2/\Upsilon) \in [6, 9]$ Mpc$^{-1}$, producing DA EDE curves with different $z_c$. 
    The residuals are taken with respect to the curve at lowest $z_c = 2.2 \times 10^5$. 
    In effect, we project that increasing $z_c$ beyond the posteriors shown in Secs.~\ref{sec:results_H0}-\ref{sec:results_LLS} does not affect our conclusions. 
    }
    \label{fig:higher_z_c}
\end{figure}

As described in our priors in Sec.~\ref{sec:priors} and in Table~\ref{tab:priors}, and indeed as found by other investigations of the sensitivity of the CMB to high redshifts \cite{Karwal:2016vyq, Linder:2010wp}, 
we find diminishing returns of searching the parameter space that permits $z_c > 10^5$. 
In Fig.~\ref{fig:higher_z_c}, we show the impact of further increasing $z_c$ on observables, while holding all other cosmological parameters constant. 
Increasing $z_c$ by over an order of magnitude has minimal impact on the CMB power spectra, all well under the cosmic variance limit of $\Delta C_\ell ^{XX}/\sigma_{\rm CV} = 1$ on the plot.  

Physically, this is unsurprising as follows. 
The redshift range shown is deep in the radiation-dominated era. 
Then, because DA EDE redshifts like radiation, simply fixing $f_\ede$ fixes the amount of dark radiation for all $z < z_c$. 
This would not be true closer to $z_{\rm eq}$, where matter density which redshifts slower than DA EDE becomes important. 
Therefore, it is unsurprising that going to higher $z_c$ has an asymptotic impact on CMB observables, as the phenomenology of the Universe at $z < z_c$ remains identical, 
resulting in the Universe simply having the same amount of dark radiation being injected earlier and earlier. 
Eventually, data would be expected to lose sensitivity to the injection time in the deep past. 
This is echoed in Fig.~\ref{fig:higher_z_c}, results that are robust to changes in $m$. 

We note that this phenomenology - the insensitivity of data to new physics introduced at asymptotically high redshifts may be present in entirely unrelated fundamental models. 
New physics with sharp transitions like in EDE \cite{Karwal:2016vyq,Poulin:2018dzj}, or step transitions like in some modified gravity, decaying dark matter models \cite{Poulin:2016nat} and redshift-dependent entropy deposits \cite{Aloni:2021eaq} only impact cosmology if data is sensitive to physics at the transition redshifts. 
If data constraints push the transition redshift to be asymptotically early, this should be read as the data disfavoring the transition.

\subsection{Why does data not prefer EDE thermal friction?}
\label{sec:results_no_EDE_solution}

Although at the background level, DA EDE can map onto the $z_c \approx z_{\rm eq}$ EDE solution to the Hubble tension \cite{Berghaus:2019cls}, it is the details of the fundamental model forming the EDE that dictate its perturbative behavior. 
Unlike the original EDE \cite{Poulin:2018cxd}, for DA EDE injection close to $z_{\rm eq}$, the fit to the CMB is significantly worsened. 

\begin{figure}
    \centering
    \includegraphics[width=0.5\textwidth]{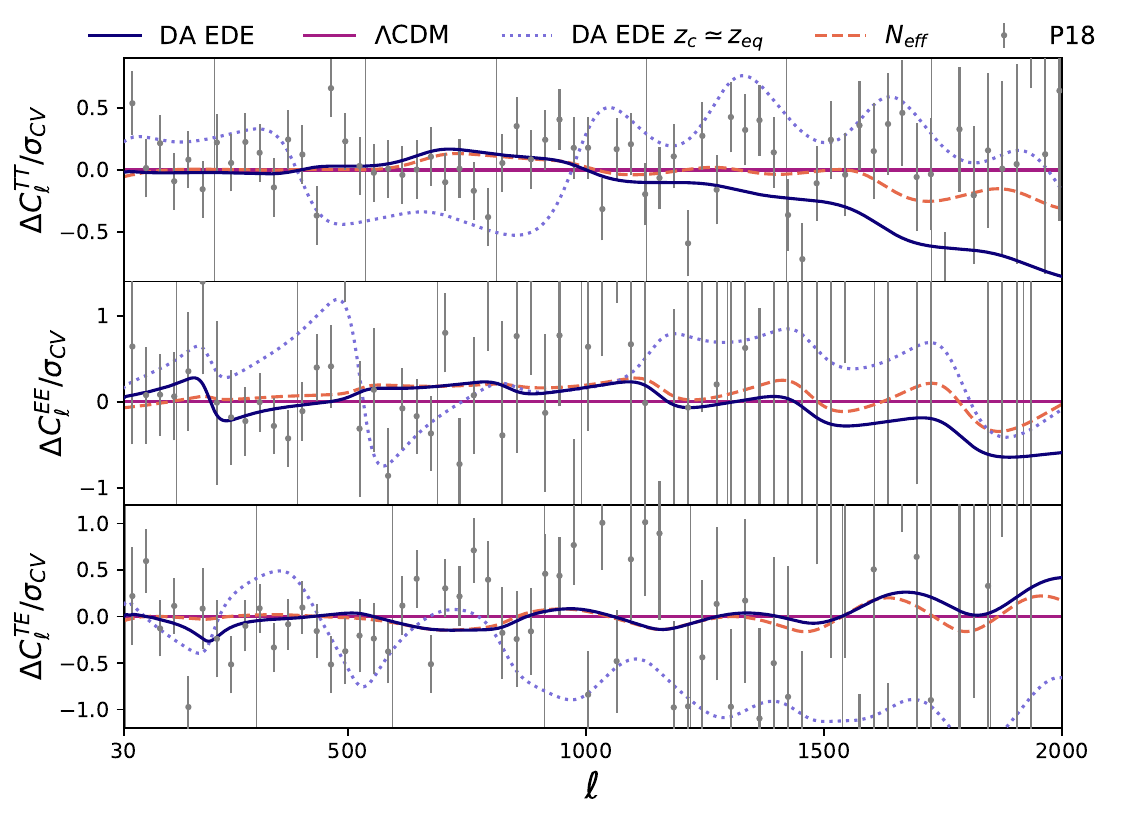}
    \caption{
    We show the CMB residuals with respect to \LCDM\ in units of cosmic variance for each cosmology at its best-fit point when fit to \textit{baseline}+$H_0$. 
    The solid curve at zero is \LCDM, $\Neff$ is in dashed orange, 
    DA EDE in solid blue, and DA EDE fixing $z_c = z_{\rm eq}$ and $f_{\ede} = 6\%$ is in dotted blue. 
    Over these, we scatter the Planck 2018 (P18) measurements of the CMB spectra. 
    Finally, the vertical lines in each plot represent the locations of the peaks in each spectrum. 
    Note that Planck 2018 nuisance parameters slightly modify these curves in their comparison with data. 
    }
    \label{fig:best_fit_cls}
\end{figure}

To further investigate the preference against an EDE-like injection, we fix $z_c = z_{\text{eq}}$ and $f_{\text{ede}} =0.06$, and search for the best-fit point in this forced EDE-like regime, by running an MCMC and then a minimizer as described in Sec.~\ref{sec:data_method} for \textit{baseline}+$H_0$. 
By fixing $f_\ede$ to the best fit for unconstrained DA EDE for the same datasets, we attempt to isolate the impact of changing the injection redshift. 
We show the CMB residuals with respect to \LCDM\ for the forced EDE best fit in Fig. \ref{fig:best_fit_cls}, along with the best-fit curves for $\Neff$ and an unconstrained DA EDE.
The incompatibility between the forced EDE curve and CMB data are apparent, most so in the first three acoustic peaks of the TT spectrum and the high-$\ell$ TE. 
These residuals cannot be absorbed by shifts in the \LCDM\ parameters, as done for other EDEs \cite{Poulin:2018cxd,Smith:2019ihp,Karwal:2021vpk}. 

We trace the source of the preference for very early $z_c$ to the impact of thermal friction terms in the DA EDE perturbation equations. 
As discussed in more detail Sec. \ref{subsec:thermal friction perturbation}, the perturbations of the dark radiation fluid dominate the impact on observables, since the scalar field quickly vanishes. 
In particular, the thermal friction terms in Eq.~\eqref{eq:rhopt1} lead to a suppression of the dark radiation density anisotropies close to the injection redshift $z_c$. 
However, for asymptotically early injection redshifts, the impact of these terms on observables is negligible. 
We conclude that data disfavor the EDE-like regime of the DA EDE model we investigate in this work due to an incompatibility with the suppression of the dark radiation anisotropies arising from thermal friction. 

\section{Conclusions}
\label{sec:conclusions}

The transformation of cosmology into a precision science has unearthed discrepancies within a \LCDM\ description of the Universe, in particular, the Hubble and large-scale structure tensions. 
Seeking to find a new concordance model of cosmology, we aim to resolve both tensions with the introduction of new physics. 

To do so, we combined favorable characteristics of two solutions to the individual tensions - early dark energy and extra radiation, into the dissipative axion model \cite{Berghaus:2019cls} (DA EDE),
which couples a scalar field to self-interacting dark radiation\footnote{In fact self-interacting radiation has been shown to outperform $\Neff$ which parameterizes extra free-streaming radiation \cite{Blinov:2020hmc, Aloni:2021eaq}.}. 
This is achieved through thermal friction acting on a scalar field, which extracts the energy density of the scalar into a dark radiation bath. 

Here, we limit ourselves to the range in which thermal friction dominates over Hubble friction around the redshift $\sim z_c$ at which the scalar thaws and donates its energy to dark radiation. 
This regime has several advantages. 
Most importantly, this addresses the primary criticism of EDE from the theory perspective - we are able to obviate fine-tuning problems related to the scalar potential. 
In this regime, the scalar is overdamped by friction and does not undergo oscillation in its potential, but instead quickly dumps all its energy density into dark radiation. Therefore there is no need for a potential in which the kinetic energy of the scalar field can dilute quickly as no sizeable component of kinetic energy is produced in this model. 
A secondary advantage of this approach is that our results become broadly applicable to all choices of potential for the scalar. 

With this set up, we calculate the perturbative behavior of the described DA EDE in synchronous gauge. As expected, observables are more sensitive to the perturbations of the dark radiation and not the fleetingly dominant scalar field.
We then constrain this model using Planck 2018 CMB, baryon acoustic oscillation and Pantheon supernova data, \tk{alone and} including either or both the local Hubble measurement and weak lensing and galaxy clustering data from DES 
to test the viability of this model in the context of the two cosmological tensions. 

For all combinations of data sets we consider, we find a preference for the injection of dark radiation to occur at redshifts $z_c > 10^4$.
We show that at these redshifts DA EDE asymptotes to a cosmology in which self-interacting dark radiation has always been part of the cosmic history.\footnote{Note that these constraints were obtained for DA EDE injected post BBN, such that the CLASS calculations of BBN are unchanged from LCDM.}
The CMB looses sensitivity to the injecting time of dark radiation at very high redshifts; we expect this loss of sensitivity to be generalizable to other redshift-dependent new physics.
We pinpoint the induced dynamics of the thermal friction in the dark radiation perturbations as the source of the preference for early injection.
Sizeable thermal friction terms suppress the anisotropies of the dark radiation, an effect that the CMB data disfavors, thus pushing the injection of the dark radiation to early times.

The principal result we highlight is when fitting to all aforementioned data. 
In DA EDE, we find both a higher $H_0$ and slightly lower $S_8$ than in the comparable model of extra massless neutrino degrees of freedom, labeled $\Neff$. 
We find a substantially higher $H_0$ than \LCDM, but also a slightly larger $S_8$. 
Ultimately, this leads us to conclude that while DA EDE is an interesting model offering novel solutions to the fine-tuning criticisms of EDE, it does not restore cosmological concordance, nor does its best fit to data possess distinct differences from extra self-interacting radiation. 

Comparing goodness-of-fits of the three models for all data sets we consider, DA EDE is comparable to \LCDM\ for both $\chi^2_{\rm CMB}$ and $\chi^2_{\rm DES}$ (see Table~\ref{tab:base_h_des_chis} and \ref{tab:full_base_h_des}), but substantially improves on $\Delta \chi^2_{H_0} = -14.4$. 
On the other hand, $\Neff$ while comparable on DES, worsens the CMB fit relative to \LCDM\ with $\Delta \chi^2_{\rm CMB} = +4.9$ with a smaller improvement to SH0ES of $\Delta \chi^2_{H_0} = -13.6$. 
Hence the species of extra radiation introduced by DA EDE offers a better fit to all data compared to $\Neff$, with a relative $\Delta \chi^2_{\rm total} = -6.15$ for two more degrees of freedom. 

We note that while these improvements are encouraging, DA EDE does not fit the CMB as well as \LCDM\ can when discrepant data sets are excluded. 
A \LCDM\ model fit to just the Planck 2018 CMB spectra, BAO measurements and Pantheon supernova data (which are consistent within a \LCDM\ description of the Universe) has $\chi^2_{\rm CMB} = 2773.2$. 
A worse fit results when \LCDM\ is fit to the discrepant data sets SH0ES and DES in addition to all the above with $\chi^2_{\rm CMB} = 2778.4$. 
While some EDE models can simultaneously accommodate the SH0ES $H_0$ yet fit the CMB with a $\chi^2_{\rm CMB} \simeq 2773.2$ or better, DA EDE does not achieve this. 
Hence, the manner in which the perturbations of our dark radiation differ from $\Neff$, specifically by not having shear perturbations, aid in fitting data better than $\Neff$, but not to the extent that we maintain the excellent fit that \LCDM\ offers the CMB. 
This is in accordance with the results of several other investigations of extra radiation as well as EDEs that dilute like radiation in the literature: 
the tensions are eased but not resolved \cite{Brinckmann:2020bcn, Agrawal:2019lmo, Poulin:2018cxd, Aghanim:2018eyx, Riess:2016jrr, Poulin:2018zxs} and the fit to data is not as good as in \LCDM \cite{Aghanim:2018eyx, Riess:2016jrr, Poulin:2018zxs}. 

In terms of parameter constrains, we find that the combination of all data sets shows preference for $f_\ede > 0$ at $>2\sigma$, with a best fit of $f_\ede = 4.7\%$ ($\Delta\Neff =0.35$) and $z_c =8 \times 10^4 $. 
This effectively corresponds to the amount of extra strongly-interacting dark radiation preferred by data.
We verify that allowing for higher $z_c$ does not impact our results, and that this result can be understood simply as the Universe always having extra dark radiation.
\tk{Lastly, comparing models using Bayesian information criteria, we find that DA EDE is strongly disfavored over \LCDM\ in all data combinations we explore, with a minimum increase of $\Delta$BIC $= 11.39$ in the \textit{baseline}$+H_0$ case. }

Discrepancies in cosmology such as the Hubble and LSS tensions may offer hints about the physics of the dark sector and a more fundamental concordance model that can succeed \LCDM. 
Although DA EDE solves the fine-tuning problems of the EDE scalar-field potential and alleviates the Hubble tension, it cannot do so while offering a good fit to data. 
Our results indicate a data-driven preference disfavoring the introduction of new physics with smoothed anisotropies, a lesson that may inform future model-building efforts aiming to resolve the Hubble tension.

\begin{acknowledgments}
We are grateful to
Gustavo Marques-Tavares, Vivian Poulin, Marco Raveri, Martin Schmaltz and Neelima Sehgal for useful discussions. 
\tk{We also thank Thomas Tram and Julien Lesgourgues for providing an early update to the CLASS cosmology code. }
We are grateful to Rouven Essig for his support in accessing computational resources, and we would like to thank the Stony Brook Research Computing and Cyberinfrastructure, and the Institute for Advanced Computational Science at Stony Brook University for access to the high-performance SeaWulf computing system, which was made possible by a $1.4$M National Science Foundation grant (\#1531492).
KB acknowledges the support of NSF Award PHY1915093. 
TK was supported by NASA ATP Grant 80NSSC18K0694 and by funds provided by the Center for Particle Cosmology at the University of Pennsylvania. 

\end{acknowledgments}

\appendix

\section{Derivation of thermal friction equations} 
\label{sec:derivations}
\subsection{Framework}
We start with decomposing the scalar field $\phi$ and the dark radiation density $\rho_{\text{dr}}$ and pressure $p_{\text{dr}}$ into a smooth background component with only time dependence, and its spatially varying perturbations
\begin{align}
    \phi(\tau,\vec{x}) &= \phi(\tau) + \delta \phi(\tau,\vec{x}) , \\
    \rho_{\text{dr}}(\tau,\vec{x}) &= \rho_{\text{dr}}(\tau) + \delta \rho_{\text{dr}}(\tau,\vec{x}), \,\,{\rm and} \\
    p_{\text{dr}}(\tau,\vec{x}) &= p_{\text{dr}}(\tau) + \delta p_{\text{dr}}(\tau,\vec{x}) .
\end{align}
We follow the notation in \cite{Ma:1995ey}, and define synchronous gauge as 
\begin{equation}
    ds^2 = a(\tau)^2 \left(-{d\tau}^2 + \left(\delta_{ij} + h_{ij}\right) dx^i dx^j \right) \,. 
\end{equation}
Here $a$ is the scale factor and $\tau$ is conformal time.
Only keeping scalar perturbations, we decompose the metric perturbations into two scalar fields $h(\tau,\Vec{x})$ and $\mu(\tau,\Vec{x})$ which correspond to the trace $h = h_{ii}$ (summing over $i$), and a traceless part $h^{||}_{ij} = \left(\partial_i \partial_j - \frac{1}{3}\delta_{ij}\nabla^2 \right) \mu(\tau,\Vec{x})$. We Fourier transform these quantities such that we are able to quantify the metric perturbations by $h(\tau,k)$, and $\eta(\tau,k)$ as 
\begin{align}
    h(\tau,\Vec{x}) &= \int d^3k e^{i \Vec{k} \cdot \Vec{x}} h(\tau,\Vec{k}) \,\,, \,\,{\rm and} \\
    \mu(\tau,\Vec{x}) &= -\int d^3k e^{i \Vec{k} \cdot \Vec{x}} \frac{1}{k^2}\left( h(\tau,\Vec{k}) + 6\eta(\tau,\Vec{k})\right) \,.
\end{align}
Similarly, we also Fourier transform the fluid and scalar field perturbations, such that $\partial_i \delta \phi = i k_i \delta \phi$, $\partial_i \delta \rho_{\text{dr}} = i k_i \delta \rho_{\text{dr}}$, and $\partial_i \delta p_{\text{dr}} = i k_i \delta p_{\text{dr}}$, where $i$ is a spatial index.

\subsection{Scalar field equations}

To derive the equation of motion of the scalar field, we compute 
\begin{equation} \label{stress}
    - \nabla_{\mu} T^{\mu 0}_{\phi} = g^{0 \alpha} \left( -\Upsilon(\rho_{\text{dr}}) {v_{\text{dr}} ^\mu} \partial_{\mu} \phi \,  \partial_{\alpha} \phi \right),
\end{equation}
where
\begin{equation}
    T^{\phi}_{\mu \nu} = \partial_\mu \phi \, \partial _\nu \phi -g_{\mu \nu} \left(\frac{1}{2} g^{\alpha \beta} \partial_{\alpha} \phi \, \partial_{\beta} \phi +V  \right),
\end{equation}
and 
\begin{align}
        -g^{0\alpha}\left( \Upsilon {v_{\text{dr}} ^\mu} \partial_{\mu} \phi  \partial_{\alpha} \phi \right) 
        = & \Upsilon a^{-3} {\phi'}^2 
        + \delta \Upsilon a^{-3} {\phi'}^2 \nonumber \\
        &+ 2 \Upsilon a^{-3} \delta \phi' \phi'  
\end{align}
up to linear order in perturbations. 
Here $v^{\mu}_{\text{dr}} = \frac{dx^\mu}{dt}$. Note that only $v^0_{\text{dr}} = \frac{1}{a}$ is non-zero at the background level. 
Multiplying equation \eqref{stress} by a factor of  $(-\frac{a^4}{\phi'})$, we arrive at 
\begin{equation} \label{eq:bg}
    \phi''(\tau) + \left(2 \mathcal{H}  +   a \Upsilon(\rho_{\text{dr}}) \right) {\phi}'(\tau) + a^2 V_\phi(\phi) = 0      
\end{equation}
for the background evolution of the scalar field, and 
\begin{eqnarray} \label{eq:pert}
    \delta \phi'' + 2 \mathcal{H} \delta \phi' + \left(k^2 +a^2 V_{\phi \phi} \right) \delta \phi  & \\ = 
    \nonumber
    -\frac{h' \phi'}{2}
    -  \Upsilon a  \delta \phi' - \frac{n}{4} \Upsilon \delta_{\text{dr}} a \phi'  
\end{eqnarray}
for its linear perturbations. In equation \ref{eq:pert} we used
\begin{equation}
    \Upsilon(\rho_{\text{dr}}) \equiv c_n \rho^{\frac{n}{4}}_{\text{dr}}   
\end{equation}
to rewrite $\delta \Upsilon = \frac{n}{4} \Upsilon \delta_{\text{dr}}$, where $\delta_{\text{dr}} = \frac{\delta \rho_{\text{dr}}}{\rho_{\text{dr}}}$.

\subsection{Dark Radiation equations}

To derive the fluid equations, we compute 
\begin{equation}
    \nabla_{\mu} T_{\text{dr}}^{\mu 0} =- g^{0\alpha}\left( \Upsilon {v_{\text{dr}} ^\mu} \partial_{\mu} \phi  \partial_{\alpha} \phi \right) \,.
\end{equation}
Replacing $\frac{\delta P_{\text{dr}}}{\delta\rho_{\text{dr}}} \equiv c_s^2$, and $\frac{p_{\text{dr}}}{\rho_{\text{dr}}} \equiv w_{\text{dr}}$, we find the fluid equation at the background level to be:
\begin{equation}
    \rho_{\text{dr}}' -3 \mathcal{H}(1+w_{\text{dr}}) \rho_{\text{dr}} = \frac{\Upsilon}{a} {\phi'}^2 \,,
\end{equation}
where we multiplied both sides by a factor of $a^2$. 

Rewriting $\delta \rho_{\text{dr}}' =\rho_{\text{dr}} \delta'_{\text{dr}} + \rho_{\text{dr}}' \frac{\delta \rho_{\text{dr}}}{\rho_{\text{dr}}}$, and plugging in the background solution for $\rho'_{\text{dr}}$, we find 
\begin{eqnarray} \label{eq:pertr}
    \delta'_{\text{dr}} 
    + 3 \mathcal H( c^2_s - w_{\text{dr}}) \delta_{\text{dr}}  
    =  \frac{(\frac{n}{4}-1)\Upsilon {\phi'}^2}{a\rho_{\text{dr}}} \delta_{\text{dr}}  \nonumber \\
    + 2\frac{\Upsilon}{a \rho_{\text{dr}}} \delta \phi' \phi' 
    - (1+w_{\text{dr}}) \left( \frac{h'}{2} 
    +\theta_{\text{dr}} \right).
\end{eqnarray}
Further defining $\theta \equiv i k^i v_i$ and computing 
\begin{equation}
    \nabla_{\mu} T_{\text{dr}}^{\mu i} =- g^{i\alpha}\left( \Upsilon {v_{\text{dr}} ^\mu} \partial_{\mu} \phi  \partial_{\alpha} \phi \right) \,,
\end{equation}
where to linear order in perturbations
\begin{equation}
    - g^{i\alpha}\left( \Upsilon {v_{\text{dr}} ^\mu} \right) =-\frac{\Upsilon}{a} i k_i \delta \phi \,,
\end{equation}
we find the velocity perturbation equation to be 
\begin{eqnarray} \label{eq:pertr2}
    & {\theta'}_{\text{dr}}+ \left(\frac{\Upsilon}{a \rho_{\text{dr}}}  \phi'^2 + \mathcal{H} (1-3w_{\text{dr}}) + \frac{w'_{\text{dr}}}{1+w_{\text{dr}}} \right){\theta}_{\text{dr}}  
    \\
    \nonumber &
   =   k^2 \frac{c^2_s}{(1+w_{\text{dr}})} \delta_{\text{dr}} + k^2 \frac{\Upsilon}{a(1+w_{\text{dr}}) \rho_{\text{dr}}} \phi' \delta \phi \,,
\end{eqnarray}
where we have multiplied both sides by a factor of $a^2ik^i$. 
In this derivation, we have assumed shear perturbations to be negligible ($T^{i \neq j}_{\text{dr}} =0$). 
This assumption is justified by self-interactions of the dark radiation being efficient in the model we are considering, which suppresses any shear and the fluid is well-described as a perfect fluid. 
We also set $w_{\text{dr}} = c^2_s = \frac{1}{3}$.
Plugging those in and taking the friction to be constant ($n = 0$), reduces Eqs. \eqref{eq:bg}, \eqref{eq:pert}, \eqref{eq:pertr}, \eqref{eq:pertr2} to the equations shown in Sec. \ref{sec:thermal friction}. 

\section{BIC calculation}
\label{app:BIC}

\tk{
The Bayesian information criteria (BIC) \cite{Trotta:2008qt} allows for model comparison with a harsh penalty on models with extra parameters as 
\begin{align}
    {\rm BIC} &\equiv -2\ln \mathcal{L}_{\rm max} + k \ln N \\
    &= -2\ln(e^{-\chi_{\rm min}^2/2}) + k \ln N \\
    &= \chi^2_{\rm min} + k\ln N \,,
\end{align}
with the model that minimizes the BIC being most preferred. 
Here $\mathcal{L}_{\rm max}$ is the maximum likelihood value and $\chi^2_{\rm min}$ the equivalent minimized goodness of fit as reported in Tables~\ref{tab:full_base}-\ref{tab:full_base_h_des}, $k$ is the total number of parameters and $N$ is the total number of data points. 
For data we consider, 
$N_{\rm CMB} = 2352$ \cite{Planck:2019nip,Planck:2018lbu}, 
$N_{\rm SNe} = 1048$ \cite{Pan-STARRS1:2017jku}, 
$N_{\rm BAO}=5$ \cite{BOSS:2016wmc,Ross:2014qpa,Beutler:2011hx}, 
$N_{\rm DES} = 30$ \cite{DES:2017myr}, 
and $N_{H_0} = 1$ \cite{Riess:2021jrx}. 
The only data sets with nuisance parameters are CMB (21) and DES (20). 
Finally, \LCDM\ has 6 parameters to which $\Neff$ adds 1 and DA EDE adds 3. 
Ultimately, the figure of interest is the $\Delta$BIC presented in Tables \ref{tab:full_base}-\ref{tab:full_base_h_des}. 
}

\section{Full MCMC posteriors}
\label{app:full_mcmc_results}

\begin{figure*}
    \centering
    \includegraphics[width=0.99\textwidth]{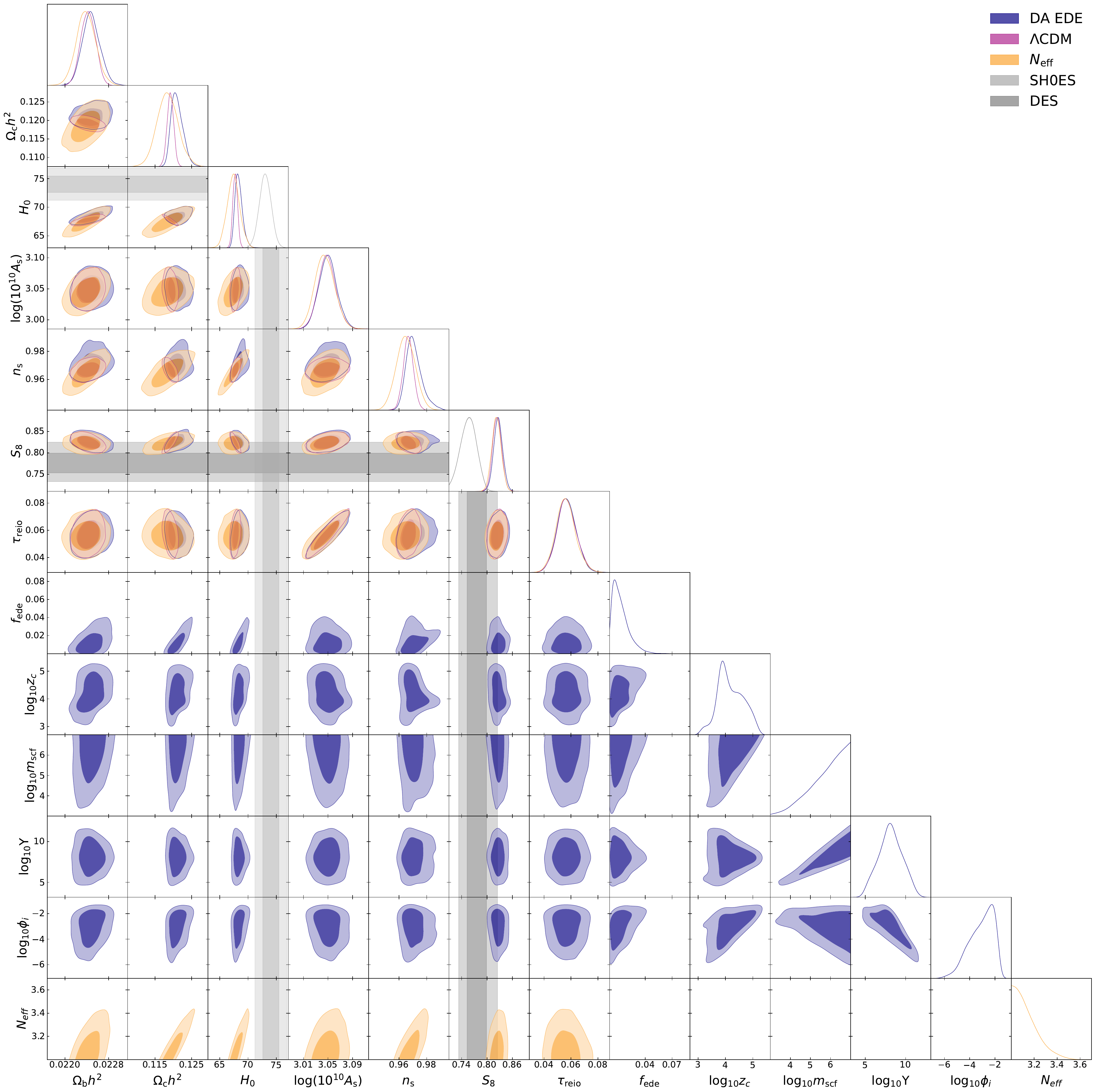}
    \caption{
    Full MCMC posteriors for all \LCDM\ and DA EDE cosmological parameters obtained by fitting to \textit{baseline} data. 
    The grey bands show the local SH0ES $H_0$ measurement, and the DES Y1 constraint on $S_8$. 
    }
    \label{fig:full_base_H0}
\end{figure*}

\begin{table*}[]
\centering
\begin{tabular}{|l|c|c|c|}
    \hline
	Parameter   &   \LCDM   &   DA EDE    &   \LCDM$+N_{\rm eff}$ \\ 
	\hline 
	\hline
$\Omega_\mathrm{b} h^2$
	 & $0.02242(0.02246)\pm 0.00013$ 
	 & $0.02248(0.02241)^{+0.00015}_{-0.00017}$ 
	 & $0.02239(0.02237)\pm 0.00018$ 
	 \\
$\Omega_\mathrm{c} h^2$
	 & $0.11920(0.11895)\pm 0.00091$ 
	 & $0.1211(0.1211)^{+0.0012}_{-0.0019}$ 
	 & $0.1185(0.1184)\pm 0.0029$ 
	 \\
$H_0$
	 & $67.73(67.86)\pm 0.42$ 
	 & $68.37(68.14)^{+0.50}_{-0.77}$ 
	 & $67.4(67.3)\pm 1.1$ 
	 \\
$\log(10^{10} A_\mathrm{s})$
	 & $3.048(3.052)\pm 0.014$ 
	 & $3.050(3.05)\pm 0.015$ 
	 & $3.045(3.043)^{+0.015}_{-0.017}$ 
	 \\
$n_\mathrm{s}$
	 & $0.9667(0.9686)\pm 0.0037$ 
	 & $0.9709(0.9751)^{+0.0042}_{-0.0066}$ 
	 & $0.9649(0.9648)\pm 0.0068$ 
	 \\
$\tau_\mathrm{reio}$
	 & $0.0569(0.0586)^{+0.0067}_{-0.0075}$ 
	 & $0.0568(0.0542)\pm 0.0072$ 
	 & $0.0562(0.0558)^{+0.0065}_{-0.0076}$ 
	 \\
$\sigma_8$
	 & $0.8102(0.8113)\pm 0.0060$ 
	 & $0.8150(0.8166)^{+0.0063}_{-0.0075}$ 
	 & $0.8075(0.8072)\pm 0.0097$ 
	 \\
$S_8$
	 & $0.8239(0.8226) \pm 0.0104$ 
	 & $0.8265(0.8308) \pm 0.0107$ 
	 & $0.8224(0.8237) \pm 0.0109$ 
	 \\
     \hline
$f_{\rm ede}$
	 & 	
	 & $0.0121(0.01)^{+0.0044}_{-0.011}$ 
	 & 	
	 \\
$\log_{10}z_c$
	 & 	
	 & $4.19(3.79)^{+0.44}_{-0.52}$ 
	 & 	
	 \\
$\log_{10}m_{\rm scf}$
	 & 	
	 & $> 5.5(5.5)$ 
	 & 	
	 \\
$\log_{10}\Upsilon$
	 & 	
	 & $8.1(8.2)\pm 1.5$ 
	 & 	
	 \\
$\phi_i$
	 & 	
	 & $0.00395(0.00034)^{+0.00089}_{-0.0056}$ 
	 & 	
	 \\
      \hline
$N_\mathrm{eff}$
	 & 	
	 & 	
	 & $3.00(2.99)\pm 0.17$ 
	 \\
     \hline
     \hline
$\chi^2_\mathrm{CMB}$
	 & $2773.2$ 
	 & $2772.2$ 
	 & $2772.7$ 
	 \\
$\chi^2_\mathrm{BAO}$
	 & $5.42$ 
	 & $5.61$ 
	 & $5.9$ 
	 \\
$\chi^2_\mathrm{SN}$
	 & $1034.92$ 
	 & $1035.01$ 
	 & $1035.1$ 
	 \\
      \hline
$\chi^2_\mathrm{total}$
	 & $3813.54$ 
	 & $3812.82$ 
	 & $3813.7$ 
      \\
 $\kb{\chi^2_\mathrm{red}}$
	 & $\kb{1.1289}$ 
	 & \kb{$1.1297$} 
	 & \kb{$1.1293$} 
      \\     
\tk{$\Delta$BIC}
	 & \tk{$0$ }
	 & \tk{$23.68$ }
	 & \tk{$8.29$ }
      \\
	 \hline
\end{tabular}
\caption{ \tk{The mean (best-fit) $\pm 1\sigma$ for each parameter for various cosmologies fit to just the \textit{baseline} data sets Planck CMB, BAO and Pantheon supernovae. } }
\label{tab:full_base}
\end{table*}

\begin{figure*}
    \centering
    \includegraphics[width=0.99\textwidth]{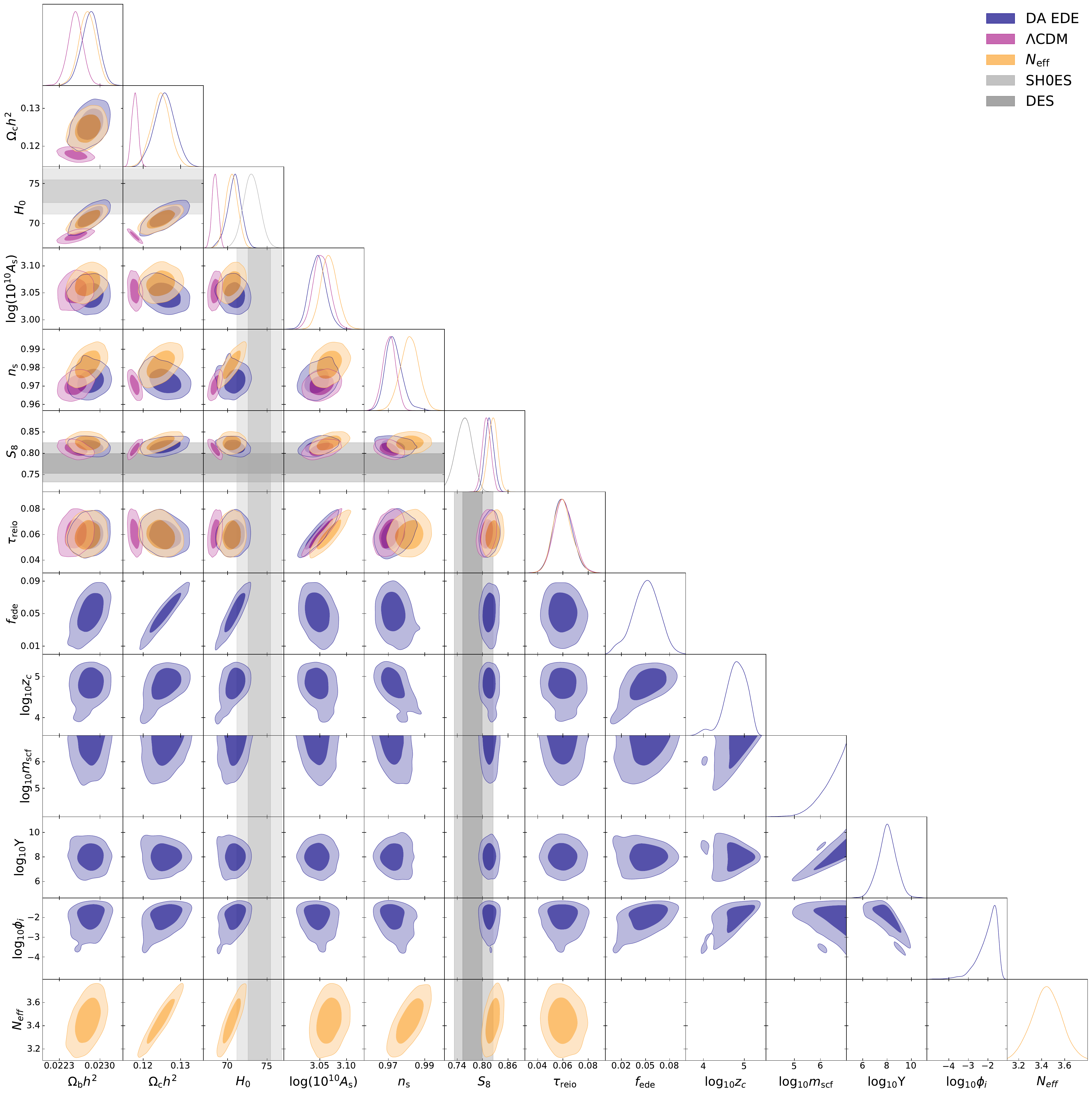}
    \caption{
    Following the same conventions as previous triangle plots, here we show the MCMC posteriors for all cosmological parameters while fitting to \textit{baseline} and the local SH0ES $H_0$ measurement. 
    }
    \label{fig:full_base_H0}
\end{figure*}

\begin{table*}[]
\centering
\begin{tabular}{|l|c|c|c|}
    \hline
	Parameter   &   \LCDM   &   DA EDE    &   \LCDM$+N_{\rm eff}$ \\ 
	\hline 
	\hline
$\Omega_\mathrm{b} h^2$
	 & $0.02258(0.0226)\pm 0.00013$ 
	 & $0.02283(0.02289)^{+0.00016}_{-0.00014}$ 
	 & $0.02278(0.02276)\pm 0.00015$ 
	 \\
$\Omega_\mathrm{c} h^2$
	 & $0.11776(0.11756)\pm 0.00085$ 
	 & $0.1257(0.1277)\pm 0.0028$ 
	 & $0.1247(0.1234)\pm 0.0025$ 
	 \\
$H_0$
	 & $68.44(68.53)\pm 0.39$ 
	 & $70.85(71.43)^{+0.93}_{-0.80}$ 
	 & $70.53(70.25)\pm 0.76$ 
	 \\
$\log(10^{10} A_\mathrm{s})$
	 & $3.053(3.059)\pm 0.015$ 
	 & $3.046(3.041)^{+0.014}_{-0.016}$ 
	 & $3.068(3.071)\pm 0.016$ 
	 \\
$n_\mathrm{s}$
	 & $0.9705(0.9716)\pm 0.0036$ 
	 & $0.9730(0.9712)^{+0.0038}_{-0.0050}$ 
	 & $0.9817(0.9812)\pm 0.0050$ 
	 \\
$\tau_\mathrm{reio}$
	 & $0.0608(0.0645)\pm 0.0078$ 
	 & $0.0600(0.0587)^{+0.0070}_{-0.0079}$ 
	 & $0.0599(0.0625)^{+0.0066}_{-0.0084}$ 
	 \\
$\sigma_8$
	 & $0.8080(0.8097)\pm 0.0064$ 
	 & $0.8197(0.8205)\pm 0.0070$ 
	 & $0.8273(0.8263)\pm 0.0088$ 
	 \\
$S_8$
	 & $0.8093(0.8095) \pm 0.0100$ 
	 & $0.8159(0.8157) \pm 0.0102$ 
	 & $0.8241(0.8228) \pm 0.0111$ 
	 \\
	 \hline
$f_{\rm ede}$
	 & 	
	 & $0.050(0.063)^{+0.018}_{-0.015}$ 
	 & 	
	 \\
$\log_{10}z_c$
	 & 	
	 & $4.79(4.96)^{+0.30}_{-0.20}$ 
	 & 	
	 \\
$\log_{10}m_{\rm scf}$
	 & 	
	 & $> 6.2(6.3)$ 
	 & 	
	 \\
$\log_{10}\Upsilon$
	 & 	
	 & $8.01(7.47)\pm 0.76$ 
	 & 	
	 \\
$\phi_i$
	 & 	
	 & $0.0155(0.031)^{+0.0059}_{-0.015}$ 
	 & 	
	 \\
	 \hline
$N_\mathrm{eff}$
	 & 	
	 & 	
	 & $3.44(3.38)\pm 0.13$ 
	 \\
     \hline
     \hline
$\chi^2_\mathrm{CMB}$
	 & $2777.5$ 
	 & $2780.3$ 
	 & $2780.0$ 
	 \\
$\chi^2_\mathrm{BAO}$
	 & $5.52$ 
	 & $6.1$ 
	 & $5.76$ 
	 \\
$\chi^2_\mathrm{SN}$
	 & $1034.736$ 
	 & $1034.746$ 
	 & $1034.739$ 
	 \\
$\chi^2_{H_0}$
	 & $18.8$ 
	 & $2.4$ 
	 & $7.2$ 
	 \\
	 \hline
$\chi^2_\mathrm{total}$
	 & $3836.56$ 
	 & $3823.55$ 
	 & $3827.7$ 
	 \\
\kb{$\chi^2_\mathrm{red}$}
	 &\kb{ $1.1354$} 
	 & \kb{$1.1326$} 
	 & \kb{$1.1331$ }
	 \\
\kb{$\Delta$BIC}
	 &\kb{ $0$} 
	 & \kb{$11.39$} 
	 & \kb{$-0.73$ }
	 \\
	 \hline
\end{tabular}
\caption{The mean (best-fit) $\pm 1\sigma$ for each parameter for various cosmologies fit to \textit{baseline} and SH0ES.} 
\label{tab:full_base_h}
\end{table*}

\begin{figure*}
    \centering
    \includegraphics[width=0.99\textwidth]{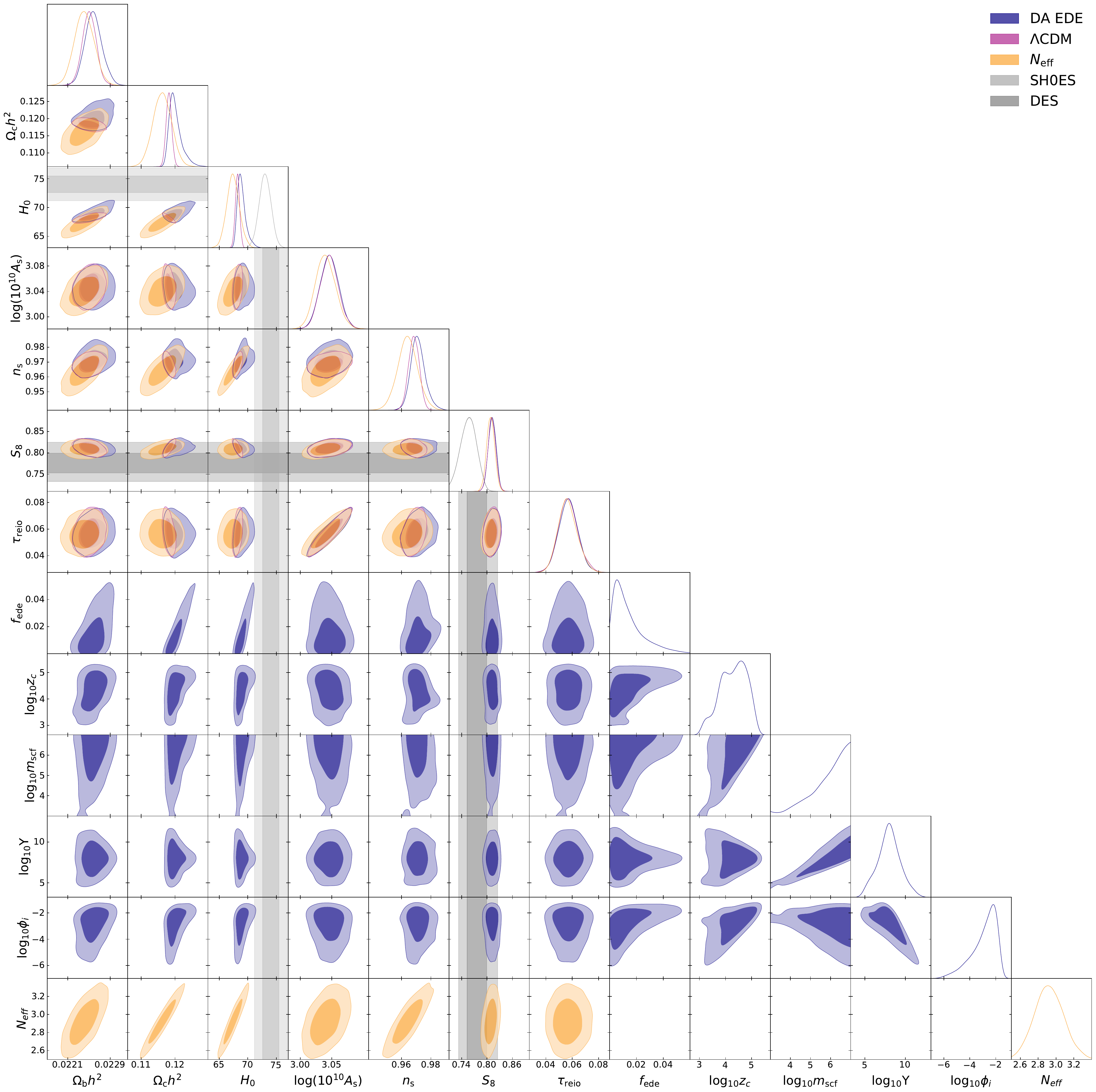}
    \caption{
    Following the same conventions as previous triangle plots, here we show the MCMC posteriors for all cosmological parameters while fitting to \textit{baseline} and DES Y1. 
    }
    \label{fig:full_base_des}
\end{figure*}

\begin{table*}[]
\centering
\begin{tabular}{|l|c|c|c|}
    \hline
	Parameter   &   \LCDM   &   DA EDE    &   \LCDM$+N_{\rm eff}$ \\ 
	\hline 
	\hline
$\Omega_\mathrm{b} h^2$
	 & $0.02251(0.02248)\pm 0.00013$ 
	 & $0.02258(0.02251)^{+0.00015}_{-0.00017}$ 
	 & $0.02241(0.02239)\pm 0.00018$ 
	 \\
$\Omega_\mathrm{c} h^2$
	 & $0.11818(0.11837)\pm 0.00083$ 
	 & $0.1201(0.1187)^{+0.0011}_{-0.0022}$ 
	 & $0.1162(0.1154)\pm 0.0027$ 
	 \\
$H_0$
	 & $68.19(68.08)\pm 0.38$ 
	 & $68.94(68.38)^{+0.46}_{-0.88}$ 
	 & $67.4(67.2)\pm 1.1$ 
	 \\
$\log(10^{10} A_\mathrm{s})$
	 & $3.046(3.043)\pm 0.015$ 
	 & $3.046(3.049)\pm 0.015$ 
	 & $3.040(3.041)\pm 0.016$ 
	 \\
$n_\mathrm{s}$
	 & $0.9684(0.9686)\pm 0.0036$ 
	 & $0.9712(0.9715)^{+0.0042}_{-0.0052}$ 
	 & $0.9642(0.9633)\pm 0.0069$ 
	 \\
$\tau_\mathrm{reio}$
	 & $0.0571(0.0548)^{+0.0068}_{-0.0078}$ 
	 & $0.0571(0.0569)\pm 0.0073$ 
	 & $0.0567(0.0585)^{+0.0065}_{-0.0079}$ 
	 \\
$\sigma_8$
	 & $0.8062(0.8058)\pm 0.0058$ 
	 & $0.8099(0.8088)\pm 0.0067$ 
	 & $0.8002(0.799)\pm 0.0095$ 
	 \\
$S_8$
	 & $0.8115(0.8128) \pm 0.0091$ 
	 & $0.8120(0.8133) \pm 0.0091$ 
	 & $0.8086(0.8078) \pm 0.0094$ 
	 \\
	 \hline
$f_{\rm ede}$
	 & 	
	 & $0.0135(0.0033)^{+0.0043}_{-0.013}$ 
	 & 	
	 \\
$\log_{10}z_c$
	 & 	
	 & $4.30(3.75)^{+0.60}_{-0.46}$ 
	 & 	
	 \\
$\log_{10}m_{\rm scf}$
	 & 	
	 & $> 5.6(6.0)$ 
	 & 	
	 \\
$\log_{10}\Upsilon$
	 & 	
	 & $8.0(9.4)\pm 1.4$ 
	 & 	
	 \\
$\phi_i$
	 & 	
	 & $0.0052(0.0)^{+0.0017}_{-0.0068}$ 
	 & 	
	 \\
	 \hline
$N_\mathrm{eff}$
	 & 	
	 & 	
	 & $2.92(2.87)\pm 0.17$ 
	 \\
     \hline
     \hline
$\chi^2_\mathrm{CMB}$
	 & $2774.1$ 
	 & $2774.7$ 
	 & $2776.0$ 
	 \\
$\chi^2_\mathrm{BAO}$
	 & $5.23$ 
	 & $5.24$ 
	 & $5.21$ 
	 \\
$\chi^2_\mathrm{SN}$
	 & $1034.82$ 
	 & $1034.776$ 
	 & $1034.86$ 
	 \\
$\chi^2_{DES}$
	 & $509.3$ 
	 & $509.4$ 
	 & $508.1$ 
	 \\
	 \hline
$\chi^2_\mathrm{total}$
	 & $4323.45$ 
	 & $4324.12$ 
	 & $4324.17$ 
	 \\
 \kb{$\chi^2_\mathrm{red}$}
	 & \kb{$1.2761$} 
	 & \kb{$1.2774$} 
	 & \kb{$1.2767$} 
	 \\
 \kb{$\Delta$BIC}
	 & \kb{$0$} 
	 & \kb{$25.1$} 
	 & \kb{$8.86$} 
	 \\
	 \hline 
\end{tabular}
\caption{The mean (best-fit) $\pm 1\sigma$ for each parameter for various cosmologies fit to \textit{baseline} and DES.}
\label{tab:full_base_des}
\end{table*}

\begin{figure*}
    \centering
    \includegraphics[width=0.99\textwidth]{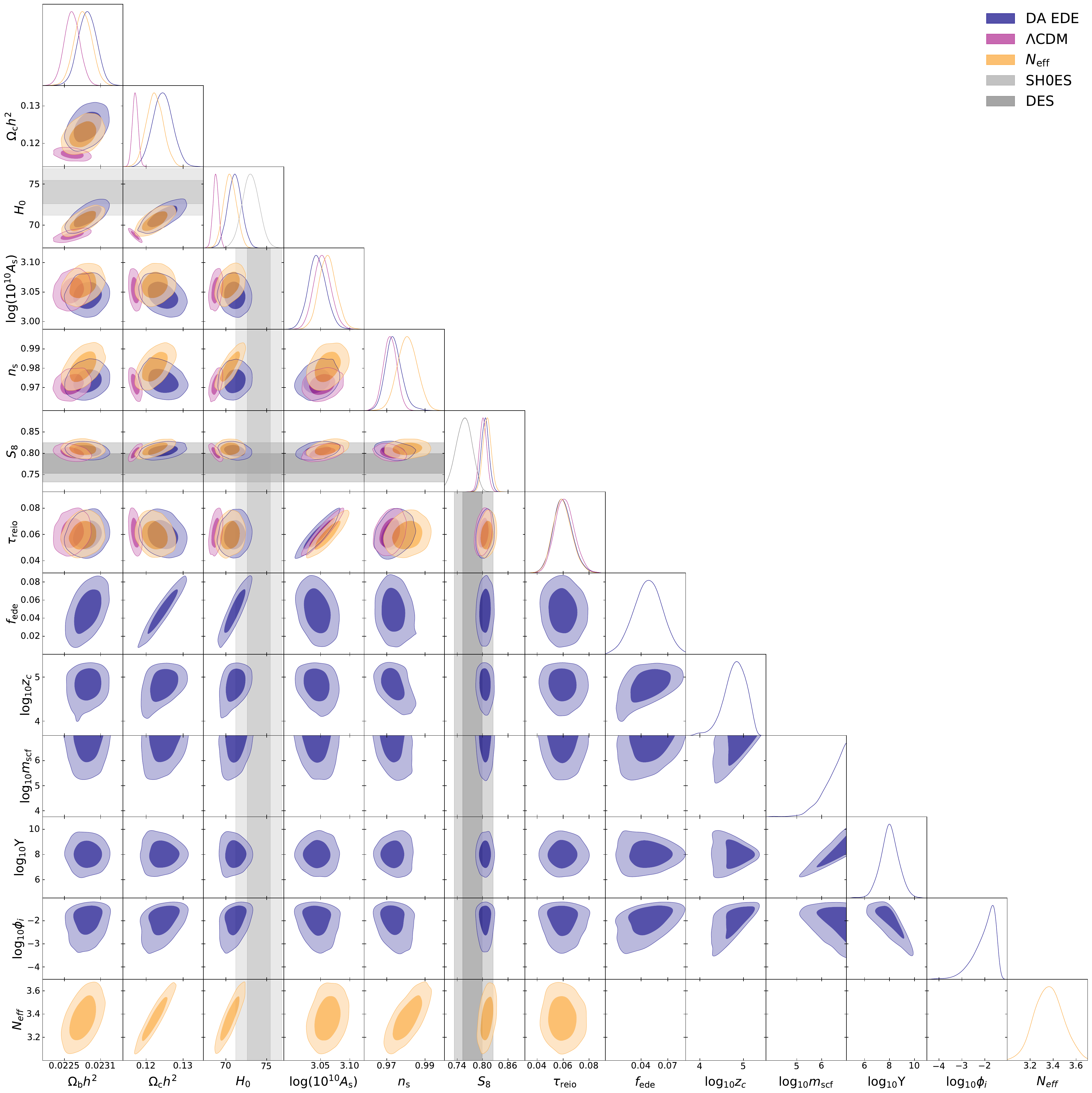}
    \caption{
    Posteriors following the same conventions as previous triangle plots, fitting to \textit{baseline}, the local SH0ES measurement of $H_0$ and DES Y1. 
    }
    \label{fig:full_base_h_des}
\end{figure*}

\begin{table*}[]
\centering
\begin{tabular}{|l|c|c|c|}
    \hline
	Parameter   &   \LCDM   &   DA EDE    &   \LCDM$+N_{\rm eff}$ \\ 
	\hline 
	\hline
$\Omega_\mathrm{b} h^2$
	 & $0.02263(0.02261)\pm 0.00013$ 
	 & $0.02288(0.02288)\pm 0.00015$ 
	 & $0.02281(0.02288)\pm 0.00015$ 
	 \\
$\Omega_\mathrm{c} h^2$
	 & $0.11705(0.11732)\pm 0.00077$ 
	 & $0.1245(0.1242)\pm 0.0027$ 
	 & $0.1224(0.1226)\pm 0.0023$ 
	 \\
$H_0$
	 & $68.76(68.63)\pm 0.36$ 
	 & $71.08(71.06)\pm 0.85$ 
	 & $70.50(70.86)\pm 0.78$ 
	 \\
$\log(10^{10} A_\mathrm{s})$
	 & $3.053(3.056)\pm 0.015$ 
	 & $3.045(3.052)\pm 0.015$ 
	 & $3.062(3.07)^{+0.014}_{-0.016}$ 
	 \\
$n_\mathrm{s}$
	 & $0.9717(0.9713)\pm 0.0035$ 
	 & $0.9736(0.9731)^{+0.0037}_{-0.0045}$ 
	 & $0.9808(0.9833)\pm 0.0051$ 
	 \\
$\tau_\mathrm{reio}$
	 & $0.0613(0.0629)^{+0.0070}_{-0.0079}$ 
	 & $0.0599(0.0641)^{+0.0068}_{-0.0078}$ 
	 & $0.0596(0.0621)^{+0.0066}_{-0.0078}$ 
	 \\
$\sigma_8$
	 & $0.8056(0.8076)\pm 0.0058$ 
	 & $0.8156(0.8179)\pm 0.0066$ 
	 & $0.8195(0.8231)\pm 0.0081$ 
	 \\
$S_8$
	 & $0.8013(0.8055) \pm 0.0087$ 
	 & $0.8058(0.8075) \pm 0.0089$ 
	 & $0.8102(0.8106) \pm 0.0096$ 
	 \\
	 \hline
$f_{\rm ede}$
	 & 	
	 & $0.048(0.047)\pm 0.016$ 
	 & 	
	 \\
$\log_{10}z_c$
	 & 	
	 & $4.81(4.91)^{+0.29}_{-0.20}$ 
	 & 	
	 \\
$\log_{10}m_{\rm scf}$
	 & 	
	 & $> 6.3(6.6)$ 
	 & 	
	 \\
$\log_{10}\Upsilon$
	 & 	
	 & $8.01(8.19)\pm 0.74$ 
	 & 	
	 \\
$\phi_i$
	 & 	
	 & $0.0152(0.0109)^{+0.0072}_{-0.015}$ 
	 & 	
	 \\
	 \hline
$N_\mathrm{eff}$
	 & 	
	 & 	
	 & $3.36(3.39)^{+0.12}_{-0.13}$ 
	 \\
\hline \hline
$\chi^2_\mathrm{CMB}$
	 & $2778.4$ 
	 & $2778.7$ 
	 & $2783.3$ 
	 \\
$\chi^2_\mathrm{BAO}$
	 & $5.7$ 
	 & $7.1$ 
	 & $7.3$ 
	 \\
$\chi^2_\mathrm{SN}$
	 & $1034.735$ 
	 & $1034.82$ 
	 & $1034.87$ 
	 \\
$\chi^2_{H_0}$
	 & $18.0$ 
	 & $3.6$ 
	 & $4.4$ 
	 \\
$\chi^2_{DES}$
	 & $508.0$ 
	 & $508.3$ 
	 & $508.8$ 
	 \\
	 \hline
$\chi^2_\mathrm{total}$
	 & $4344.84$ 
	 & $4332.52$ 
	 & $4338.67$ 
	 \\
\kb{$\chi^2_\mathrm{red}$}
	 & \kb{$1.2820$ }
	 & \kb{$1.2795$} 
	 & \kb{$1.2806$} 
	 \\  
\kb{$\Delta$BIC}
	 & \tk{$0$}
	 & \tk{$12.11$}
	 & \tk{$1.97$}
	 \\  
	 \hline
\end{tabular}
\caption{The mean (best-fit) $\pm 1\sigma$ for each parameter for various cosmologies fit to \textit{baseline}, SH0ES and DES.}
\label{tab:full_base_h_des}
\end{table*}

\bibliography{EDE}

\end{document}